\documentclass[aps,prx,superscriptaddress,twocolumn,longbibliography,floatfix]{revtex4-2}
\usepackage{units}
\usepackage{mathtools} 
\usepackage{amsmath}
\usepackage{amsthm}
\usepackage{amssymb}
\usepackage{graphicx}
\usepackage{color}
\usepackage{xcolor}
\usepackage{apptools}
\usepackage{appendix}

\definecolor{myurlcolor}{rgb}{0,0,0.7}
\definecolor{myrefcolor}{rgb}{0.1,0,0.9}

\usepackage[
	breaklinks,
	pdftex,
	colorlinks=true, 
	linkcolor=myrefcolor,
	citecolor=myurlcolor,
	urlcolor=myurlcolor
]{hyperref}

\usepackage{orcidlink}

\usepackage[linesnumbered, ruled,vlined]{algorithm2e}

\graphicspath{{./images/},{./imagesAppendix/}}

\renewcommand{\eqref}[1]{Eq.~(\ref{#1})} %

\def\app#1#2{%
  \mathrel{%
    \setbox0=\hbox{$#1\sim$}%
    \setbox2=\hbox{%
      \rlap{\hbox{$#1\propto$}}%
      \lower1.1\ht0\box0%
    }%
    \raise0.25\ht2\box2%
  }%
}

\ifx\proof\undefined
\newenvironment{proof}[1][\protect\proofname]{\par
	\normalfont\topsep6\p@\@plus6\p@\relax
	\trivlist
	\itemindent\parindent
	\item[\hskip\labelsep\scshape #1]\ignorespaces
}{%
	\endtrivlist\@endpefalse
}
\providecommand{\proofname}{Proof}
\fi

\newcommand{\bra}[1]{\langle #1|}
\newcommand{\ket}[1]{|#1 \rangle}
\makeatother

\newcommand{\idg}[1]{{\bfseries #1)}}

\providecommand{\factname}{Fact}

\providecommand{\claimname}{Claim}
\providecommand{\lemmaname}{Lemma}
\providecommand{\definitionname}{Definition}
\providecommand{\corollaryname}{Corollary}

\definecolor{THc}{rgb}{0.9,0.3,0.2}

\newcommand{\subfigimg}[3][,]{%
	\setbox1=\hbox{\includegraphics[#1]{#3}}%
	\leavevmode\rlap{\usebox1}%
	\rlap{\hspace*{2pt}\raisebox{\dimexpr\ht1-0.5\baselineskip}{{\bfseries \large\textsf{#2}}}}%
	\phantom{\usebox1}%
}

\newcommand{\sectionMain}[1]{
\let\oldaddcontentsline\addcontentsline%
\renewcommand{\addcontentsline}[3]{}%
\section{#1}
\let\addcontentsline\oldaddcontentsline
}

\newcommand{\eq}[1]{\hyperref[eq:#1]{(\ref*{eq:#1})}}

\AtAppendix{\counterwithin{theorem}{section}}
\AtAppendix{\counterwithin{lemma}{section}}
\AtAppendix{\counterwithin{fact}{section}}
\AtAppendix{\counterwithin{definition}{section}}

\allowdisplaybreaks

\begin{document}

\title{Rise and fall of nonstabilizerness via random measurements}

\author{Annarita Scocco~\orcidlink{0000-0002-6920-5843}}
\email{annarita.scocco@unina.it}
\affiliation{Scuola Superiore Meridionale, Largo San Marcellino 10, I-80138 Napoli, Italy}

\author{Wai-Keong Mok~\orcidlink{0000-0002-1920-5407}}
\email{darielmok@caltech.edu}
\affiliation{Institute for Quantum Information and Matter, California Institute of Technology, Pasadena, CA 91125, USA}

\author{Leandro Aolita}
\affiliation{Quantum Research Center, Technology Innovation Institute, Abu Dhabi, UAE}

\author{Mario Collura~\orcidlink{0000-0003-2615-8140}}
\email{mcollura@sissa.it}
\affiliation{International School for Advanced Studies (SISSA), via Bonomea 265, 34136 Trieste, Italy}
\affiliation{INFN Sezione di Trieste, 34136 Trieste, Italy}

\author{Tobias Haug~\orcidlink{0000-0003-2707-9962}}
\email{tobias.haug@u.nus.edu}
\affiliation{Quantum Research Center, Technology Innovation Institute, Abu Dhabi, UAE}

\begin{abstract}
We investigate the dynamics of nonstabilizerness - also known as `magic' - in monitored quantum circuits composed of random Clifford unitaries and local projective measurements. For measurements in the computational basis, we derive an analytical model for dynamics of the stabilizer nullity, showing that it decays in quantized steps and requires exponentially many measurements to vanish, which reveals the strong protection through Clifford scrambling. 
On the other hand, for measurements performed in rotated non-Clifford bases, measurements can both create and destroy nonstabilizerness. Here, the dynamics leads to a steady-state with non-trivial nonstabilizerness, independent of the initial state. We find that Haar-random states equilibrate in constant time, whereas stabilizer states exhibit linear-in-size relaxation time.
While the stabilizer nullity is insensitive to the rotation angle, Stabilizer Rényi Entropies expose a richer structure in their dynamics. 
Our results uncover sharp distinctions between coarse and fine-grained nonstabilizerness diagnostics and demonstrate how measurements can both suppress and sustain quantum computational resources.
\end{abstract}

\maketitle

 \let\oldaddcontentsline\addcontentsline%
\renewcommand{\addcontentsline}[3]{}%

\section{Introduction}
In the quest for quantum advantage, significant effort has been devoted to identifying and formalizing the resources that enable quantum computational power, often through the framework of resource theories. 
While entanglement is a key signature of \emph{quantumness}, it is not sufficient for quantum computational advantage. This is evidenced by the existence of states that can be efficiently simulated on classical computers despite being highly entangled, notably the class of stabilizer states. 
These are generated by Clifford operations~\cite{gottesman1997stabilizer,gottesman1998heisenberg,aaronson2004improved}, built from Hadamard, phase, and controlled-NOT gates, as well as by projective measurements in the computational basis. 
Stabilizers do not unlock universal quantum computation on their own, as they have to be augmented with non-Clifford operators - such as $T$-gates~\cite{nielsen2011quantum,leone2024learningtdoped} - for quantum advantage~\cite{howard2014contextuality,veitch2014resource}. They constitute the main factor determining the runtime of quantum computers~\cite{campbell2017roads} and are exponentially hard to simulate on quantum computers~\cite{bravyi2019simulation,bu2019efficient,leone2021quantum}.

The amount of non-Clifford operations necessary to construct generic quantum states can be quantified through the resource theory of nonstabilizerness or `magic'~\cite{kitaev2003fault,bravyi2005universal,chitambar2019quantum,liu2022many,howard2017application,veitch2014resource,liu2022many,horodecki2013quantumness,veitch2012negative,heinrich2019robustness,seddon2021quantifying,campbell2010bound,howard2014contextuality}.
Nonstabilizerness measures include, among others, the stabilizer nullity~\cite{beverland2020lower,jiang2021lower} and stabilizer R\'enyi entropies (SREs)~\cite{leone2022stabilizer}. The latter are particularly powerful: They be analytically treated~\cite{lopez2024exact,turkeshi2023pauli}, numerically efficiently computed for different classes of states~\cite{haug2023stabilizer,haug2022quantifying,tarabunga2023many,lami2023quantum,passarelli2024nonstabilizerness,tarabunga2025efficientmutualmagicmagic,frau2024non} and experimentally measured~\cite{haug2022scalable,oliviero2022measuring,haug2023efficient,haug2025efficient}. They offer a lens for understanding a wide variety of phenomena in many-body physics~\cite{gu2024doped,rattacaso2023stabilizer,oliviero2022magic,odavic2022complexity} and quantum chaos~\cite{passarelli2025chaos,leone2021quantum,leone2023nonstabilizerness,magni2025quantum,jasser2025stabilizerentropyentanglementcomplexity}.

The interplay between quantum resources and quantum operations is now a rich area of exploration~\cite{howard2017application,leone2022stabilizer,tirrito2023quantifying,haug2024probing,gu2025magicinduced}. Clifford gates — free operations in the resource theory of magic — do not increase magic, yet they create a significant amount of entanglement. 
In contrast, projective measurements tend to suppress entanglement and destroy quantum information~\cite{benzion2020disentangling}. However, applying high-entanglement random unitaries can effectively spread the information throughout the Hilbert space, hiding it from local measurements~\cite{choi2020quantum,fidkowski2021dynamical}. 
This principle is the key feature used in quantum error correction to protect logical information within many redundant physical qubits~\cite{gottesman1997stabilizer,brown2013short}. 
Furthermore, systems combining unitary Clifford dynamics with random projective measurements have been shown to exhibit a measurement-induced phase transition in entanglement entropy~\cite{skinner2019measurement,li2019measurement,li2018quantumzeno,tarabunga2024magic,bejan2023dynamical,fux2023entanglementmagic,russomanno2025efficient} depending on the value of the measurement rate: frequent measurements suppress entanglement, leading to area-law scaling and loss of information~\cite{gullans2020dynamical}, while infrequent measurements allow unitary dynamics to dominate, resulting in volume-law growth and error suppression~\cite{choi2020quantum}. 

Although much attention has been focused on the entanglement entropy~\cite{oliviero2021transitions,true2022transitionsin,jian2020measurement,lunt2021measurement,coppola2022growth,legal2024entanglement,lirasolanilla2024multipartite,zabalo2020critical}, increasing interest has turned to nonstabilizerness in monitored quantum circuits, exploring its dynamics under measurements. 
Nonstabilizerness has been studied in different settings such as Clifford+T circuits~\cite{bejan2023dynamical,fux2023entanglementmagic}, random quantum circuits~\cite{leone2023phase,zhou2020single,turkeshi2025magic,gu2025magicinduced} or monitored Gaussian fermions~\cite{tirrito2025magic}.
In Ref.~\cite{paviglianiti2024estimating}, the authors investigated the decay of magic under sequences of local Cliffords and measurements, finding that the decay of magic can be characterized by two phases, depending on the rate of measurements. 
Further, Refs.~\cite{niroula2023phase,trigueros2025} studied the injection of magical noise channels within a random Clifford basis followed by measurements of most of the qubits, in analogy to quantum error correction under coherent noise, which is shown to remove magic below a critical noise rate. 
However, past works concentrated on the regime where the number of measurements is extensive with system size, while the role of individual measurements in the dynamics has not been well understood. 

Here, we study how random measurements create and destroy magic. We study circuits composed of global random Clifford unitaries combined with single-qubit measurements in the computational basis. We construct an analytical model to describe the dynamics of nullity, observing that it takes exponential time to decay, then extending it to the SRE. 
We then study a modified circuit where we rotate the single-qubit measurements into a non-Clifford basis, characterized by an angle parameter $\theta_M$. In this setting, measurements can both create and destroy magic and the system converges to a long-time steady-state with non-trivial magic. 
The time to convergence depends on the initial state: states with high nullity require a constant time to relax, whereas low-nullity states require a linear time in the qubit number. 
For the nullity, the steady-state is solely determined by the system size and independent of $\theta_M$. In contrast, the SRE steady-state scales as $\theta_M^2$ for $\theta_M\ll1$.
Our work shows that magic can persist and thrive even under measurements.

The manuscript is organized as follows: In Sec.~\ref{sec:nonstabilizerness} we review the magic resource measures of stabilizer nullity and stabilizer Rényi entropy (SRE). 
In Sec.~\ref{sec:protocol} we explain our circuit and measurement protocol. 
In Sec.~\ref{sec:analytical_model} we provide analytic models describing how magic evolves via non-magic measurements, while in Sec.~\ref{sec:analytical_magic} we extend our model to measurements that can induce magic. 
In Sec.~\ref{sec:numerical}, we show numerical simulations of our protocol for both nullity and SRE for different states and comparing the numerics with analytic results. 
Finally, in Sec.~\ref{sec:discussion} we discuss our results and draw our conclusions. 

\begin{figure*}[htbp]
	\centering	
 	\raisebox{0\height}{\subfigimg[width=0.44\textwidth]{a}{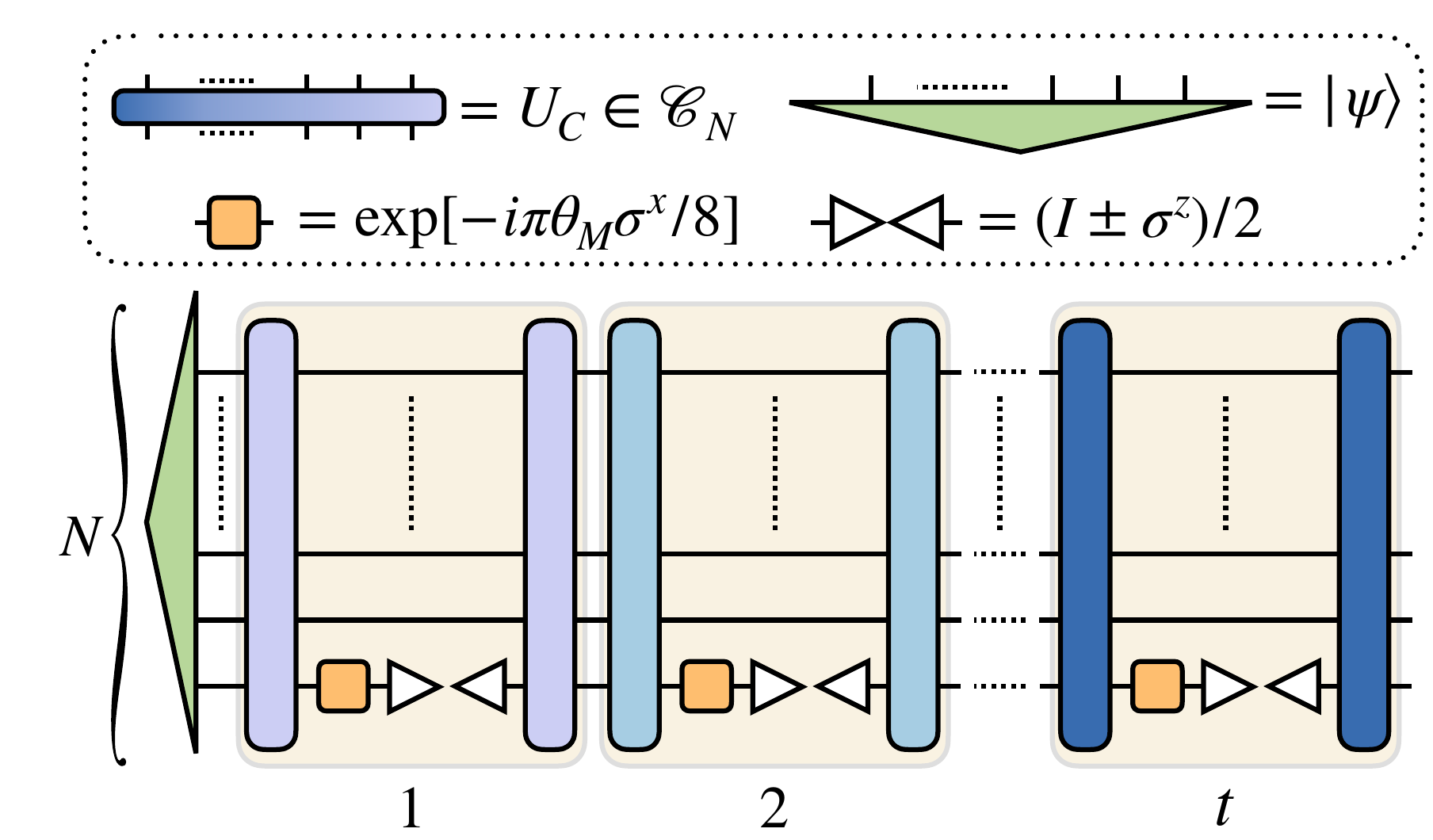}}
    \subfigimg[width=0.27\textwidth]{b}{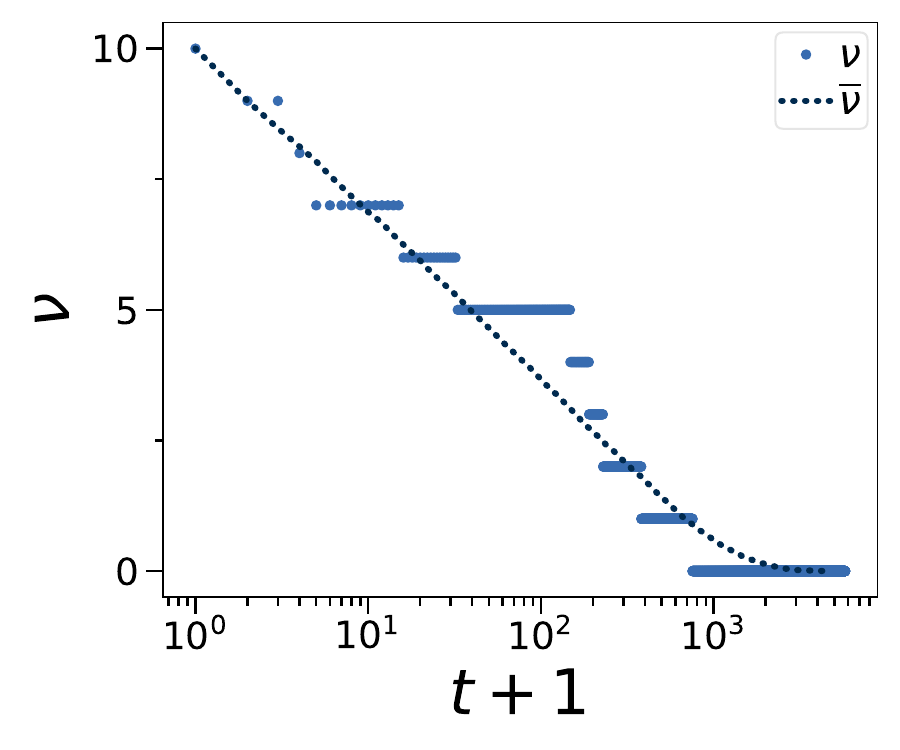}\hfill
    \subfigimg[width=0.27\textwidth]{c}{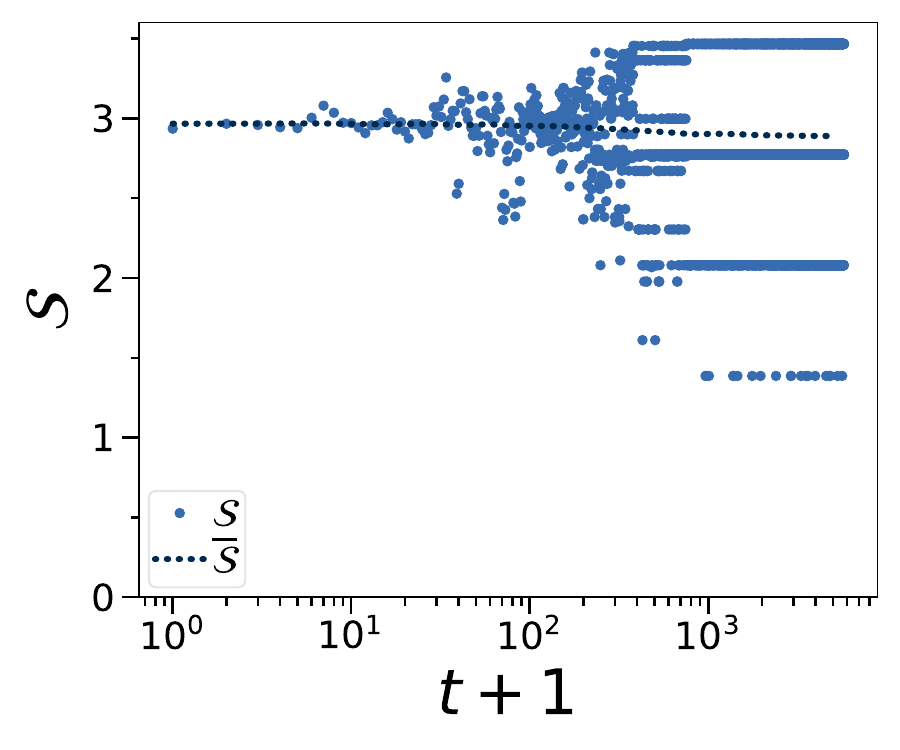}
    \caption{
    \idg{a} Circuit with random measurements in (rotated) Clifford basis. Starting from initial state $\ket{\psi}$, we apply $t$ rounds of randomly chosen global Clifford unitaries $U_\text{C}$, single-qubit rotation $R_x(\theta_\text{M})$ with angle $\theta_\text{M}\in[0,1]$, projection onto single-qubit computational basis state $\ket{b}$ sampled according to its its measurement probability $p_s$, and inverse Clifford $U_\text{C}^\dagger$. For $\theta_\text{M}=0$, circuit is completely Clifford, while it induces magic for $\theta_\text{M}>0$.
    \idg{b} Representative trajectory of stabilizer nullity $\nu$ for different measurement steps $t$ and $\theta_\text{M}=0$ for $N = 10$, together with the average $\overline{\nu}$ over $N_\text{r} \approx 1000$ trajectories in dotted line. We start from initial Haar random state, and each measurement either leaves $\nu$ invariant or decreases it, yielding a characteristic step-wise decay towards stabilizer states, where the probability for a step to occur decreases exponentially with $t$. 
    \idg{c} Entanglement as measured by von-Neumann entropy $\mathcal{S}$ for half-bipartition against $t$ for same representative trajectory. The average entanglement $\mathcal{S}$ barely changes with $t$, however individual trajectories of $\mathcal{S}$ fluctuate strongly with increasing $t$. 
    Approaching low magic, and at the end of the dynamics when the state can be described as a random stabilizer, $\mathcal{S}$ can only acquire particular values, between which the trajectory jumps rapidly. For the representative trajectory, for $t \geq 10^3$ we see that $\nu = 0$, and average $\overline{\mathcal{S}}$ corresponds to the entanglement entropy of a random stabilizer state. 
	}
    \label{fig:circuit}
\end{figure*}

\section{Nonstabilizerness}\label{sec:nonstabilizerness}

Let us consider $N-$qubit Pauli operators $ P = \bigotimes_{i=1}^N \sigma^\alpha_i$ which are tensor products of single-qubit Pauli operators $\sigma^x,\,\sigma^y$ and $\sigma^z$ and the single-qubit identity $ \sigma^0=I_1 $.
We consider $\mathcal{P}_N$ as the set of all unsigned $N$-qubit Pauli operators $P\in \mathcal{P}_N$, that is the Pauli group modulo global phases $\{\pm1, \pm i \}$, with cardinality $\vert \mathcal{P}_N \vert =4^N$. We denote $\mathcal{G}_{S}(\ket{\psi})$ as the stabilizer group of a given state $\ket{\psi}$, i.e the set of commuting Pauli operators with 
\begin{equation}
P\ket{\psi}=\pm\ket{\psi}\,,\forall P\in \mathcal{G}_{S}\,.
\end{equation}
Stabilizer states are pure quantum states where the stabilizer group has maximal size with $\mathcal{G}_{S}=2^N$.
Alternatively, any stabilizer state can be generated by applying on the computational basis state $\ket{0}^{\otimes N}$ the unitary gates of the Clifford group $U_\text{C} \in\mathcal{C}_N$, generated by the Hadamard gate, the CNOT gate and the $\pi/4$ phase gate. 
Notably, such Clifford unitaries $U_\text{C}$ map Pauli strings $P$ into Pauli strings as $P'=U_\text{C}PU_\text{C}^\dagger$. 
Using the so-called tableau formalism~\cite{gottesman1997stabilizer,aaronson2004improved}, one can store the generators of the stabilizer group in a $N \times 2N$ tableau, allowing for efficient Clifford unitary operations at cost $\mathcal{O}(N^2)$

Of course not all states are stabilizer states, and indeed nonstabilizerness provides a way to quantify how far a given quantum state is from the set of stabilizer states. Various measures have been introduced to capture different aspects of this resource~\cite{heinrich2019robustness,wang2019quantifying,haug2022scalable,bravyi2019simulation}. In the remainder of this section, we focus on two such quantities that are central to this work: the stabilizer nullity and the $\alpha-$Stabilizer Rényi Entropies.

The definition of the stabilizer nullity $\nu$ of a state~\cite{beverland2020lower,jiang2021lower} is based on the number of independent Pauli operators that stabilize it, formally
\begin{equation}
    \nu(\ket{\psi}) = N - \log_2 \vert \mathcal{G}_S (\ket{\psi})\vert\,,
\end{equation}
where $\vert \mathcal{G}_S (\ket{\psi}) \vert $ is the cardinality of the stabilizer group. According to this definition, for a general pure state $\ket{\psi}$ its nullity is such that $\nu(\ket{\psi})\leq N$, while $\nu(\ket{\psi}) = 0$ for pure stabilizer states.
$\nu(\ket{\psi})$ represents a genuine magic monotone, as it is nonincreasing under stabilizer operations, such as Clifford unitaries or measurements of Pauli operators. However it exhibits sensitivity to small perturbations: even infinitesimal non-stabilizer components in the state can reduce the number of exact Pauli stabilizers, resulting in a discontinuous jump in nullity. Despite this, its operational meaning and simplicity make it a useful and interpretable diagnostic of nonstabilizerness.

The second measure of magic we refer to are the $\alpha$-SREs~\cite{leone2022stabilizer}. They quantify the $\alpha$-th moments of the distribution of Pauli expectation values
\begin{equation}
    \mathcal{M}_{\alpha}(\ket{\psi}) = \frac{1}{1-\alpha} \ln \left( \sum_{P\in\mathcal{P}_N} 2^{-N} | \bra \psi P \ket{\psi} |^{2 \alpha}\right)\,.
\end{equation}
$\alpha-$SREs can be interpreted as the $\alpha$-R\'{e}nyi entropy of the classical probability distribution $\Xi_P  = 2^{-N}| \bra \psi P \ket{\psi} |^{2}$ since $\Xi_P \geq 0$ and $\sum_{P \in \mathcal{P}_N} \Xi_P = 1$ for any $\ket{\psi}$.  While it is not a monotone under Clifford operations for $\alpha < 2$ and not a strong monotone for any $\alpha$~\cite{haug2023stabilizer}, it is a monotone for pure states for $\alpha \geq 2$~\cite{leone2024stabilizer}. 
Note that the nullity can be used to bound the SREs as $\mathcal{M}_{\alpha}(\ket{\psi})\leq \nu(\ket{\psi}) \ln 2$.

\section{Protocol}\label{sec:protocol}

We study the evolution of magic in monitored circuits measured in random Clifford basis, which is sketched in Fig.~\ref{fig:circuit}a. 
We consider a monitored circuit where we apply a random $N$-qubit Clifford unitary $ U_\text{C}\in\mathcal{C}_N $, followed by a single-qubit rotation in the $x$-basis acting on the first qubit
\begin{equation}
    R_x(\theta_M)=\exp(-i\theta_M \pi/8\sigma^x_1)\,.
\end{equation}
Note that, due to the Clifford unitary randomizing the basis, we would get the equivalent dynamics if we choose any other qubit.
We then measure the same qubit in the computational basis $\sigma_1^z$, record the measurement outcome $\in \{\pm1\}$. Note that here measurement and rotation are orthogonal, so nonstabilizerness can remain after measurement, in contrast to the case where rotation and measurement in the same basis as studied in Ref.~\cite{bejan2023dynamical}.

Then, we apply the inverse Clifford unitary $U_\text{C}^\dagger$ on the full state. We repeat this process $t$ times, where we sample a random Clifford unitary $U_\text{C}$ at every step. 
Operationally, this process can be described by projection-valued measures (PVM)
\begin{equation}
    \Pi_{\pm}(\theta_M) = \frac{1}{2} \bigg[ \cos \frac{\theta_M \pi}{8}(I_N \pm  P^z) -i \sin \frac{\theta_M \pi}{8}(P^x \mp i P^y) \bigg],
    \label{eq:generalized_clifford}
\end{equation}
where $P^{\gamma} \equiv U_C^\dagger \sigma^\gamma_1 U_C$ is a Pauli operator rotated into a random Clifford basis with $\gamma\in\{x,y,z\}$, and $\Pi_{b}$ is a generalized measure operator~\cite{nielsen2011quantum} satisfying $\sum_{b=\pm}\Pi_{b}^{\dag}\Pi_{b} = I_N$. 
For each step, the probability of measuring the outcome $\pm1$ in the state $|\psi\rangle$ is given by $p_{\pm} = \langle \psi | \Pi_{\pm}^{\dag} \Pi_{\pm}| \psi \rangle$.  
After each measurement, we normalize the post-measurement state as $\ket{\psi_{\pm}} = \frac{1}{\sqrt{p_{\pm}}} \Pi_{\pm} \ket{\psi}$. 

For $\theta_M=0$, the process is a Clifford operation and does not induce any magic. In this case,~\eqref{eq:generalized_clifford} simplifies to a projector of rank $2^{N-1}$
\begin{equation}\label{eq:cliffordMeas}
    \Pi_{\pm}(\theta_M=0)=\frac{1}{2}(I_N \pm P)\,,
\end{equation}
with random Pauli $P\in\mathcal{P}_N/\{I\}$, and the $z$ superscript has been discarded for simplicity.
In contrast, for $\theta_M>0$, the process can induce magic due to the single-qubit rotation $R_x(\theta_M)$ before the measurement, where the induced magic is maximal for $\theta_M=1$.

To give a glimpse of the effect of such measurement protocol, we show a representative trajectory and the average over $N_{\text{r}}$ trajectories for $\theta_M=0$ in Fig.~\ref{fig:circuit}b,c, respectively the stabilizer nullity and the von Neumann entropy, defined as $\mathcal{S}(\rho_A)=\text{tr}(\rho_A\log \rho_A)$ with bipartite state $\rho_A=\text{tr}_A(\ket{\psi}\bra{\psi})$ over bipartition $A$. In the two panels we show the evolution of the two quantities starting from the same initial Haar-random state according to the same random unitaries.
In Fig.~\ref{fig:circuit}b, the nullity $\nu$ starts from its maximal value and decays at random times in a discrete stair case, yielding an average which takes an exponential number of measurements to vanish. 
In contrast, in Fig.~\ref{fig:circuit}c the average von Neumann entanglement entropy $\overline{\mathcal{S}}$ barely changes on average in time $t$. The trajectory shows initially small fluctuations which become more pronounced and discrete for large time $t$, rapidly jumping between different values. At exponential times, the state becomes a random stabilizer state ($\nu=0$), for which $\mathcal{S}$ can only assume a linear number of different values~\cite{dahlsten2005exact,toth2005entanglement,fattal2004entanglement,nahum2017quantum}. Thus, magic uncovers a complementary picture on the dynamics in random monitored circuits, which we proceed to study in more detail in the following sections.

We test our protocols on four different initial states: random states $\ket{\psi}_\text{Haar}$ drawn from the Haar measure~\cite{mele2024introduction}, tensor products of $T$-states $\ket{T}^{\otimes N}$, where $\ket{T}\propto \ket{0}+e^{-\pi/4}\ket{1}$, states $\ket{\psi}_{\text{GUE}}  = \exp(-iH_{\text{GUE}}t)\ket{0}^{\otimes N}$ evolved with random Hamiltonian $H_{\text{GUE}}$ drawn from the Gaussian unitary ensemble with normalized spectrum $\in[-2,2]$~\cite{cotler2017chaos,haug2024probing} where we choose small $t=0.1$, and finally the computational basis state $\ket{0}^{\otimes N}$. 

\section{Non-magic measurements}\label{sec:analytical_model}

\subsection{Analytical model}
We now present an analytical model of how measurements affect magic by describing the dynamics as a Markov chain, where we first consider $\theta_M=0$, while $\theta_M>0$ is deferred to Sec.~\ref{sec:analytical_magic}. 
We start with an initial state $\ket{\psi_\nu}$ with nullity $\nu$, then apply a random Clifford unitary, measure in the $z$ basis, then apply the inverse of the Clifford unitary, which corresponds to~\eqref{eq:cliffordMeas}.  
One can see from~\eqref{eq:cliffordMeas} that there $\vert  \mathcal{P}_N/\{I_N\}\vert=4^N-1$ possible choices of basis, from which we draw randomly due to the random Clifford transformation. At each round of measurement, $\nu$ either decreases, or remains unchanged. 
Any state $\ket{\psi_\nu}$ with nullity $\nu$ can be written as~\cite{paviglianiti2024estimating} 
\begin{equation}
\ket{\psi_\nu}=U_\text{C}\ket{0}^{\otimes N-\nu}\ket{M}\,,
\end{equation}
where we have Clifford unitary $U_\text{C}$ and some $\nu$-qubit magic state $\ket{M}$.
Now, let us consider the state after measuring in Pauli basis $P$
\begin{equation}
    \begin{aligned}
    \Pi_\pm\ket{\psi_\nu}&\propto(I_N \pm P)U_\text{C}\ket{0}^{\otimes N-\nu}\ket{M} \\
     &= \frac{1}{2}U_\text{C}(I_N\pm P')\ket{0}^{\otimes N-\nu}\ket{M}\,, 
     \end{aligned}
\end{equation}
where we have the transformed Pauli operator $P'=U_\text{C}^\dagger P U_\text{C}$. 
The measurement $\Pi_\pm$ can affect stabilizer nullity $\nu$ in three distinct ways depending on the chosen Pauli $P'$~\cite{paviglianiti2024estimating}:  

$(i)$ First, let us consider a Pauli basis $P'=P_{N-\nu}\otimes I_\nu$, where $P_{N-\nu}\in\mathcal{P}_{N-\nu}/\{I_{N-\nu}\}$ acts on the first $N-\nu$ qubits, while $I_{\nu}$ is the identity acting on the last $\nu$ qubits. Then, we can neglect any prefactors and the initial Clifford $U_\text{C}$ as they do not affect the measurement.
Further, 
\begin{equation}
    (I_N\pm P_{N-\nu}\otimes I_\nu)\ket{0}^{\otimes N-\nu}\ket{M}
\end{equation}
acts trivially on the magic part $\ket{M}$ of the state, and only induces a Clifford transformation which leaves $\nu$ invariant, i.e. $\nu\rightarrow\nu$. The case $(i)$ has $(4^{N-\nu}-1)$ possible Paulis $P'$.

$(ii)$ As second case, let us consider $P'=\{I,\sigma^z\}^{\otimes N-\nu}\otimes P_\nu$, where we have a tensor product of identity and $\sigma^z$ Pauli operators on the first $N-\nu$ qubits, and $P_{\nu}\in\mathcal{P_{\nu}}/\{I_\nu\}$. Let us assume that we have $\sigma^z$ on the first $N-\nu$ qubits, while all other cases follow similarly.
Applying the measurement gives us then
\begin{equation}
    \begin{aligned}
    (I_N\pm {\sigma^z}^{\otimes N-\nu}\otimes P_\nu)\ket{0}^{\otimes N-\nu}\ket{M}\\
    =(I_N\pm I_{N-\nu}\otimes P_\nu)\ket{0}^{\otimes N-\nu}\ket{M}\,.
    \end{aligned}
\end{equation}
This is equivalent to measuring a qubit on magic state $\ket{M}$, which reduces the stabilizer nullity $\nu\rightarrow\nu-1$. Note that here we neglect the probability of the nullity decreasing by more than $1$, as such events are extremely unlikely. 
Case $(ii)$ has $2^{N-\nu}(4^{\nu}-1)$ possible Pauli operators.

$(iii)$ Finally, we have $P'=\{I,\sigma^z,\sigma^x,\sigma^y\}^{\otimes N-\nu}\otimes P_\nu$, where for the first $N-\nu$ qubits we have a tensor product of identity, $\sigma^z$, $\sigma^x$ and $\sigma^y$ Pauli operators, where at least one Pauli operator is either $\sigma^x$ or $\sigma^y$, and $P_\nu \in\mathcal{P}_\nu/\{I_\nu\}$. 
For simplicity, let us assume we have $\sigma^x$ on the first $N-\nu$ qubits, while all other cases follow similarly. Then,
\begin{equation}
    \begin{aligned}
    &(I_N\pm {\sigma^x}^{\otimes N-\nu}\otimes P_\nu)\ket{0}^{\otimes N-\nu}\ket{M}\\
    =&(\ket{0}^{\otimes N-\nu}\ket{M}
    \pm I_{N-\nu}\otimes P_\nu)\ket{1}^{\otimes N-\nu}\ket{M}\,.
    \end{aligned}
\end{equation}
Now, we add the control-Pauli $\text{c}P_\nu$ (by using $\text{c}P_\nu^\dagger\text{c}P_\nu=I$), where the control acts on one of the first $N-\nu$ qubits. Note that $\text{c}P_\nu$  is a Clifford unitary. This gives us
\begin{equation}
    \begin{aligned}
    &(I_N\pm {\sigma^x}^{\otimes N-\nu}\otimes P_\nu)\ket{0}^{\otimes N-\nu}\ket{M} \\
    = &\text{c}P_\nu^\dagger \text{c}P_\nu(\ket{0}^{\otimes N-\nu}\ket{M}
    \pm I_{N-\nu}\otimes P_\nu)\ket{1}^{\otimes N-\nu}\ket{M} \\
    = &\text{c}P_\nu^\dagger (\ket{0}^{\otimes N-\nu}\ket{M}
    \pm \ket{1}^{\otimes N-\nu}\ket{M})\\=
    &\text{c}P_\nu^\dagger\ket{+}^{\otimes N-\nu}\ket{M}\,.
    \end{aligned}
\end{equation}
Note that the final state has unchanged nullity i.e. $\nu\rightarrow \nu$.
Case $(iii)$ has $(4^{N-\nu}-2^{N-\nu})(4^{\nu}-1)$ possible Paulis.
One can easily verify that $(i)$ to $(iii)$ cover all $P\in\mathcal{P}_N/\{I\}$ with in total $4^N-1$ Pauli operators.

Now, after one timestep, the nullity can only decrease for case $(ii)$ with probability
\begin{equation}\label{eq:nullitydecay}
    \text{Pr}_\text{z}(\nu \to \nu -1) = %
    \frac{2^{N}}{4^{N} - 1}(2^\nu-2^{-\nu})
    \,,
\end{equation}
where for $\nu\gg1$ we have $\text{Pr}_\text{z}(\nu \to \nu -1) \approx 2^{\nu-N}$, i.e. exponentially small with decreasing $\nu$.
The nullity remains unchanged with probability $\text{Pr}_\text{z}(\nu \to \nu) = 1 - \text{Pr}_\text{z}(\nu \to \nu - 1)$. 

Our model now describes, in agreement with what derived in Ref.~\cite{gu2025magicinduced}, a Markov chain for the random variable $\nu$, with the update rule of the probability distribution $\rho(\nu)$ for each timestep $t$ given by
\begin{equation}\label{eq:markov}
    \rho(\nu) \to \sum_{\nu^\prime} \rho(\nu^\prime) \text{Pr}_\text{z}(\nu^\prime \to \nu)\,.
\end{equation}
Motivated by this, we now derive an approximate model for the dynamics of $\nu$. To simplify the analytical calculations, let us define the non-stabilized dimension $y$ as
\begin{equation}
    y(\nu) = 2^\nu\,,
\end{equation} and rewrite~\eqref{eq:nullitydecay} as
\begin{equation}
    \text{Pr}_\text{z}(y \to y/2) = A_N \left(y - \frac{1}{y}\right),
\end{equation}
where
\begin{equation}
    A_N = \frac{2^N}{4^N - 1}.
    \label{eq:an}
\end{equation}
Then, each step of the Markov chain changes the value of $\overline{y}$ as
\begin{equation}
    \overline{y} \to \overline{y} - \frac{A_N}{2} \left( \overline{y^2} - 1 \right),
\end{equation}
where the overbar indicates the statistical average over all realizations. Let us assume that the probability distribution of $y$ is sharply concentrated about the mean value $\overline{y}$, such that $\overline{y^2} \approx \overline{y}^2$. Then, we can construct an approximate continuous-time model
\begin{equation}
\label{eq:approx_model_nomagic}
    \frac{d\overline{y}}{dt} \approx -\frac{A_N}{2} \left(\overline{y}^2 - 1\right).
\end{equation}
In this model, the number of measurement rounds $t$ is approximated by a continuous variable, which can be interpreted as the evolution time of the system. Solving this differential equation yields
\begin{equation}
    \label{eq:approx_model_nomagic_solution}
    \overline{y}(t) \approx 1 - \frac{2b}{e^{A_N t} + b}\,,
\end{equation}
where
\begin{equation}
    b = \frac{1 - \overline{y}(0)}{1 + \overline{y}(0)}\,.
    \label{eq:b}
\end{equation}
For large $N$ and $\overline{y}(0) \gg 1$, we have $A_N \approx 2^{-N}$ and $b \approx -1$, which gives
\begin{equation}
\label{eq:approx_nullity_nomagic}
    \overline{y}(t) \approx 1 + \frac{2}{e^{t/2^{N}} - 1} \approx \coth\left(\frac{t}{2^{N}}\right)\,. 
\end{equation}
Note that $\overline{\nu}(t) \leq \log_2 \overline{y}(t)$ from Jensen's inequality. However, since we assumed that the distribution of $y$ is sharply concentrated in deriving~\eqref{eq:approx_model_nomagic}, we can approximate $\overline{\nu}(t) \approx \log_2 \overline{y}(t)$. Evaluating~\eqref{eq:approx_nullity_nomagic} for the regimes $t \ll 2^N$ and $t \gg 2^N$, yields the leading-order behavior
\begin{equation}
\overline{y}(t) \approx \begin{cases}
    \frac{2^{N}}{t}, \quad &t \ll 2^{N} \\
    1 + 2 e^{-t/2^N}, \quad &t \gg 2^{N}\,.
\end{cases}
\end{equation}
Equivalently, we have
\begin{equation}
    \overline{\nu}(t) \approx \begin{cases}
    N - \log_2 t, \quad &t \ll 2^{N} \\
    \frac{2}{\ln 2} e^{-t/2^{N}}, \quad &t \gg 2^{N}\,.
\end{cases}
\end{equation}
This implies that the stabilizer nullity (on average) decreases very slowly with the number of measurement rounds, up to an exponential timescale $t \sim 2^{N}$. At late times $t \gg 2^{N}$, the stabilizer nullity decreases exponentially to zero, where the decay factor $2^{-N}$ is exponentially small. We stress that these analytical results are meant to provide qualitative insights on the dynamics of the stabilizer nullity, and we do not expect this to exactly reproduce the discrete-time dynamics generated by the Markov chain due to the approximations used.

\subsection{Numerical results}

We now perform numerical simulations of our circuit and compare the results to our Markov chain model.
We first focus on the stabilizer nullity $\nu$. The probability that the nullity decreases with measurement is exponentially small in $\nu$, as derived in~\eqref{eq:nullitydecay}. In Fig.~\ref{fig:circuit}b we plot one representative trajectory. We find that the first measurement always induces $\nu \to \nu -1$, and then $\nu$ decreases exactly $N$ times in steps of 1, creating plateaus whose length increases exponentially with time, on average. The average $\overline{\nu}$ takes an exponential time to decay which is plotted as dotted line.

\begin{figure}[t!]
	\centering	
 	\subfigimg[width=0.3\textwidth]{}{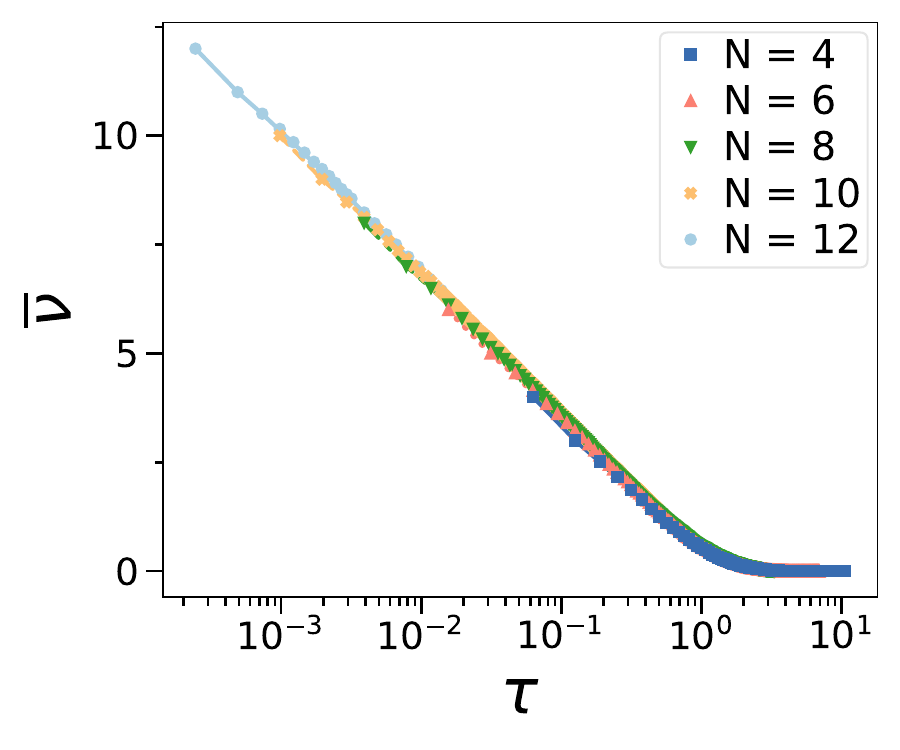}
    \caption{ 
    Average nullity $\overline{\nu} $ against $\tau$, where the rescaled time for each curve is $\tau = A_N(t+1)$, $A_N$ as defined in~\eqref{eq:an}. 
    The points represent the numerical simulation of the full quantum model via~\eqref{eq:generalized_clifford}, while the lines are a  simulation of our Markov chain model as given by~\eqref{eq:markov}.
	}
    \label{fig:nullity}
\end{figure}

We apply the measurement protocol to three different initial state types with initial maximal nullity $\overline{\nu}(0) \equiv \overline{\nu}_0 = N$, as listed in Sec.~\ref{sec:protocol}. Note that here we do not study the initial computational basis state $\ket{0}^{\otimes N}$, as our protocol leaves stabilizer states trivially unchanged. 
The agreement between the exact numerical simulation and a simulation of our Markov chain model of \eqref{eq:markov} is shown in Fig.~\ref{fig:nullity}. 
The average $\overline{\nu}$ over $N_{\text{r}} \approx 1000$ trajectories behaves in the same way regardless of the initial state type. The results depicted are obtained starting from initial Haar-random states, but the other two states yield completely equivalent results. 
In particular, different points correspond to the numerical simulation of Haar-random states for different sizes $N$, and lines with matching colors correspond to the dynamics simulated by the Markov chain rule of~\eqref{eq:markov} with the corresponding transition probabilities given by~\eqref{eq:nullitydecay}. 
Each curve is plotted as a function of time rescaled by its size-dependent timescale $A_N$ as defined in~\eqref{eq:an}, so that $t \to A_N t$ (we call $\tau = A_N(t+1)$ to show what happens at $t = 0$ in a semi-log scale). As we can see, the model correctly predicts the average decay towards random stabilizer states with $\overline{\nu} = 0$, and all curves collapse on top of each other. We also find that the analytic solution to the differential equation in~\eqref{eq:approx_model_nomagic_solution} agrees well with the exact numerical simulations, which we show in detail in Appendix~\ref{sec:comparison}.

\begin{figure*}[t!]
	\centering	
    \subfigimg[width=0.31\textwidth]{a}{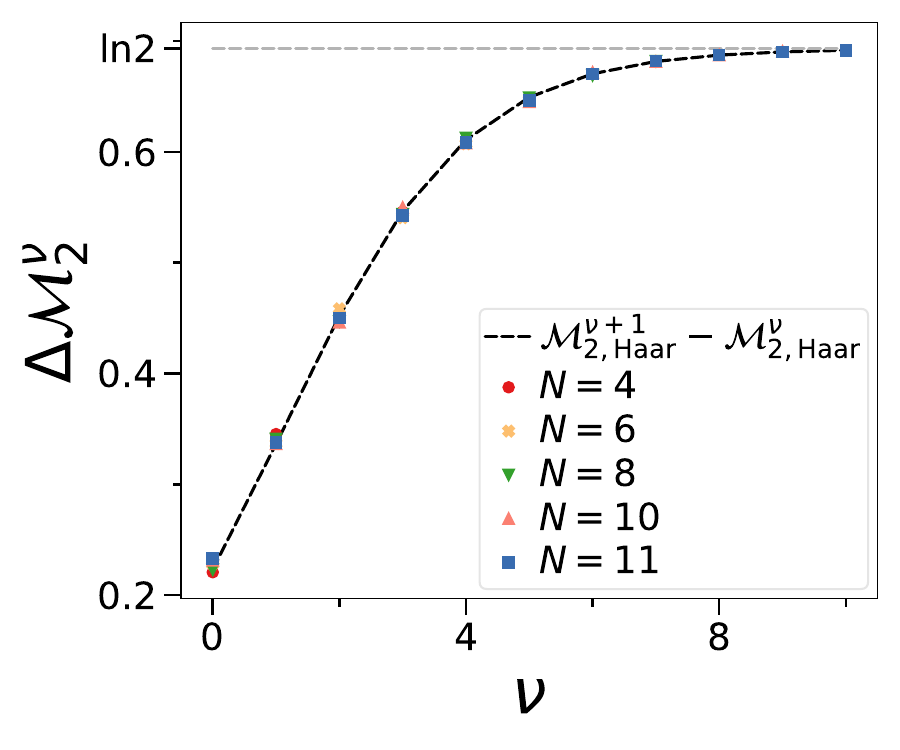}\hfill
    \subfigimg[width=0.31\textwidth]{b}{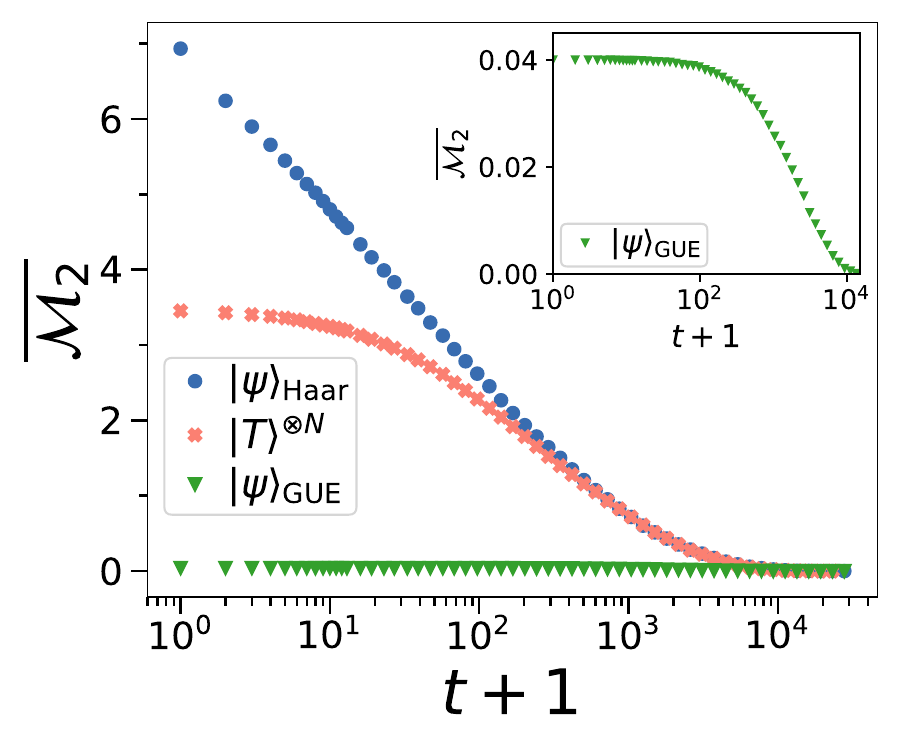}\hfill
    \subfigimg[width=0.31\textwidth]{c}{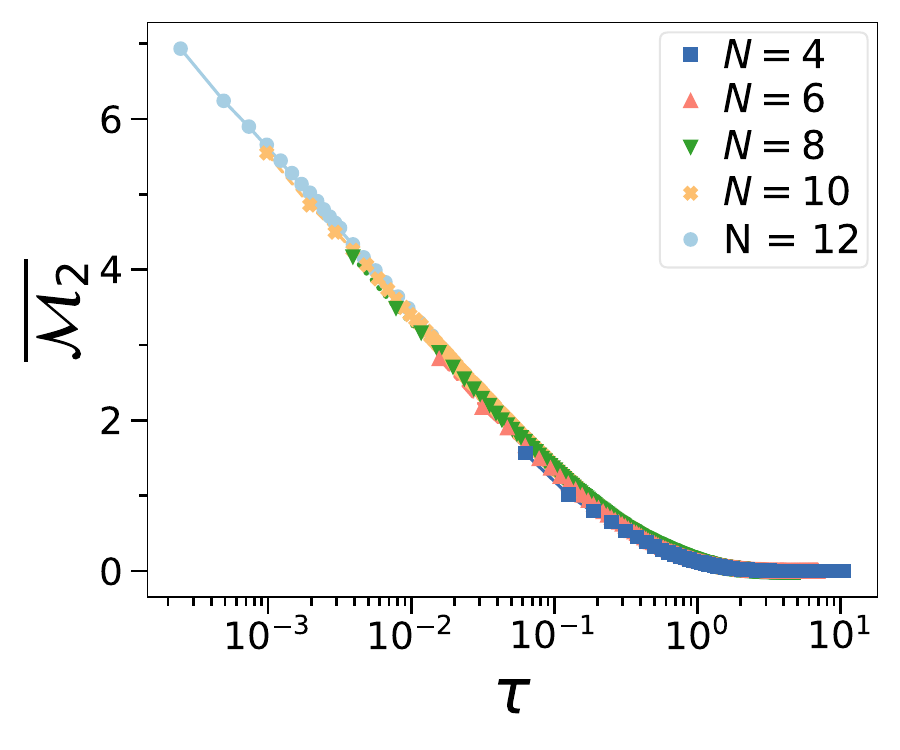}
    \caption{
    \idg{a} Change in SRE $\Delta \mathcal{M}_2^{\nu}$ at the $\nu$th measurement that yields a change in magic, averaged over $N_{\text{r}} \approx 1000$ trajectories, for different $N$ and for $\theta_M = 0$.
    The dashed line is $\overline{\mathcal{M}}_\text{Haar}^{\nu+1}  - \overline{\mathcal{M}}_\text{Haar}^{\nu}$, where $\mathcal{M}_\text{Haar}^{N}$ is the SRE of Haar random state of $N$ qubits as defined in~\eqref{eq:avgsre2_avgnullity}. 
    \idg{b} Average SRE $\overline{\mathcal{M}}_2$ over $N_{\text{r}} \approx 1000$ trajectories vs measurement step $t$ for three different initial states, $\ket{\psi}_{\text{Haar}}$, $\ket{T}^{\otimes N}$ and $\ket{\psi}_{\text{GUE}}$ for $\theta_M = 0$. Inset: Average magic $\overline{\mathcal{M}}_2$ for $\ket{\psi}_{\text{GUE}}$ with different axis range. 
    \idg{c} Dots show numerical $\overline{\mathcal{M}}_2$ against rescaled $\tau = A_N (t+1)$ for initial Haar-random states averaged over $N_{\text{r}} \approx 1000$ trajectories where $A_N\approx 2^{-N}$. Solid lines show $\overline{\mathcal{M}}_2$ obtained using nullity from analytic prediction of the Markov chain in~\eqref{eq:approx_model_nomagic_solution} inserted into SRE of Haar-random states of~\eqref{eq:avgsre2_avgnullity}.
	}
    \label{fig:sre2_theta_0}
\end{figure*}

We now describe the dynamics of the SRE $\mathcal{M}_2$ in our circuit. Similar to $\nu$, we observe discrete changes in SRE with $t$, decreasing $N$ times until it vanishes. However, for the SRE, the difference between consecutive plateaus is not constant, but can change; we show representative trajectories in Appendix~\ref{sec:single_traj}. 
Further, unlike stabilizer nullity $\nu$, the dynamics depends on the initial state. We now discuss in detail what happens for Haar-random states. We can characterize the distribution of differences in SRE between two consecutive plateaus using the following ansatz: 
After $N - \nu$ reductions in nullity, the remaining state can be written as $\nu$-qubit Haar-random state $\ket{\psi}_\text{Haar}^{\nu} $~\cite{paviglianiti2024estimating} combined with $(N - \nu)$-qubit stabilizer state in a random Clifford basis $U_\text{C}$ 
\begin{equation}\label{eq:statehaarmeas}
\ket{\psi_\nu} = U_\text{C}  \ket{\psi}_\text{Haar}^{\nu} \ket{0}^{\otimes N-\nu} \,,
\end{equation} 
where its average magic is given by~\cite{leone2022stabilizer}
\begin{equation}
    \mathcal{M}_{2,\text{Haar}}(\nu) = \ln\bigg(\frac{3 + 2^{\nu}}{4}\bigg)\,.
    \label{eq:avgsre2_avgnullity}
\end{equation}
We find that the change in SRE from the step with nullity $\nu+1$ to nullity $\nu$, which we define as $\Delta \mathcal{M}_{2}^\nu$, is on average
\begin{equation}
\Delta \mathcal{M}_{2,\text{Haar}}^\nu = \mathcal{M}_{2,\text{Haar}}^{\nu+1}  - \mathcal{M}_{2,\text{Haar}}^{\nu}\,.
\end{equation}
In particular, looking at Fig.~\ref{fig:sre2_theta_0}a, we can see that at the beginning of the evolution ($\nu\gg1$), the average difference between two consecutive plateaus is almost constant as $\Delta \mathcal{M}_2^\nu \to \ln 2$, which is predicted by~\eqref{eq:avgsre2_avgnullity}, while it becomes smaller for $\nu\approx 0$ due to finite-size effects.

As previously stated, the SRE is initial-state dependent. In general, also for $\ket{T}^{\otimes N}$ and $\ket{\psi}_{\text{GUE}}$ states, $\theta_M = 0$ measurements can either reduce SRE or keep it constant, with rare exceptions better explained in Appendix.~\ref{sec:magic_increase}. 
In Fig.~\ref{fig:sre2_theta_0}b, we then show how the average SRE $\overline{\mathcal{M}_2}$ evolves with this measurement protocol for different initial states with initial maximal nullity $\nu_0 = N$.
Despite all three still taking an exponential number of measurements to vanish, they behave differently for most part of the dynamics. In particular, the SRE for $\ket{T}^{\otimes N}$ states, after a certain number of measurements, overlaps with the SRE of $\ket{\psi}_{\text{Haar}}$ states. For the $\ket{\psi}_{\text{GUE}}$ states, $\overline{\mathcal{M}_2}$ is much lower than the one of the other two states as shown in the inset in Fig.~\ref{fig:sre2_theta_0}(b) with a different range on the $y$-axis, overlapping with the other two at the very end of the dynamics. 

We can now show that the model derived for the decay of nullity can also accurately capture that of the SRE for Haar-random states.
Indeed, for such states, the~\eqref{eq:avgsre2_avgnullity} holds and maps the nullity $\nu \to \mathcal{M}_2$ at all times. 
In Fig.~\ref{fig:sre2_theta_0}c, we show for Haar random states the average SRE $\overline{\mathcal{M}}_2$ over $N_r \sim 1000$ trajectories as different points of different colors representing increasing $N$. 
The solid lines of the same color represent the mapping through~\eqref{eq:avgsre2_avgnullity} of the $\nu$ as predicted by the Markov chain rule in~\eqref{eq:approx_model_nomagic_solution}. Time is rescaled as $ t \to A_N t$ (we again call $\tau = A_N(t+1)$) consistently with the scaling observed for stabilizer nullity as shown in Fig.~\ref{fig:nullity}, the different curves collapse onto a single universal one. 
The SRE of both $\ket{T}^{\otimes}$ and $\ket{\psi}_{\text{GUE}}$ states does not respect such functional relation, as further explained in Appendix~\ref{sec:relations_nullity_states}. Despite no analytical predictions on them can be made using this model, we can still use this relation as an effective model to phenomenologically describe SREs for different ensembles (in Appendix~\ref{sec:comparison}).

\section{Measurement in magic basis}\label{sec:analytical_magic}

\subsection{Analytic model}

Next, we develop a model for our circuit with $\theta_\text{M}>0$, i.e. the random measurements can induce magic via the rotation $R_x(\theta_\text{M})$ before the measurement.
First, we compute the probability that the rotation $R_x(\theta_\text{M})$ changes the nullity $\nu$.
Here, we can assume the rotation 
\begin{align*}
    R_P(\theta_M)&=\exp(-i\theta_M \pi/8P)\\
    &=\cos(\theta_M\pi/8)I-i\sin(\theta_M\pi/8)P
\end{align*}
acts in a random Pauli basis $P=U_\text{C}\sigma^x U_\text{C}^\dagger$ due to random Clifford unitary $U_\text{C}$.
We can write the initial state for nullity $\nu$ as $\ket{\psi_\nu}$ with Clifford unitary $U_\text{C}$ applied to a tensor product of stabilizer state $\ket{0}^{\otimes N-\nu}$ and some magic state of $\nu$ qubits $\ket{M}$~\cite{paviglianiti2024estimating}
\begin{equation}
\ket{\psi_\nu}=U_\text{C}\ket{0}^{\otimes N-\nu}\ket{M}\,.
\end{equation}
Now, the rotation $R_P(\theta_\text{M})$ does not increase $\nu$ when $P$ acts trivially on $\ket{0}^{\otimes N-\nu}$.
These are Paulis of the form $P\in\{I,\sigma^z\}^{\otimes {N-\nu}}\otimes P_\nu$, where $P_\nu \in\mathcal{P}_\nu$ and $P\neq I_N$. The corresponding probability is given by
\begin{equation}
\text{Pr}_R(\nu \rightarrow \nu)=\frac{2^{N+\nu}-1}{4^N-1}\,.
\label{eq:pr_r_nu_nu}
\end{equation}
The probability that the nullity increases is then given by
\begin{equation}
\text{Pr}_R(\nu \rightarrow \nu+1)=1-\text{Pr}_R(\nu \rightarrow \nu)=\frac{4^N-2^{N+\nu}}{4^N-1}\,.
\label{eq:pr_r_nu_increase}
\end{equation}
Here, we neglected irrelevant fringe cases: For example, $\nu$ could also potentially decrease due to the rotation on $\ket{M}$, which however is extremely unlikely.
Note that the nullity only measures whether stabilizers are broken (even by slightest amounts), and thus is practically independent of rotation angle $\theta_\text{M}$ as long as it is non-Clifford, i.e. $\theta_\text{M}\neq 0$.

After the rotation, the projective measurement is applied. 
Assuming sufficient entanglement, the measurement is nearly independent of the rotation and only depends on the nullity (after the rotation).  For measurement, the same probabilities as in the previous section (with $\theta_M=0$) apply.
Thus, we can simply multiply the probabilities of change in $\nu$ for rotation and measurement, giving us 
\begin{equation}
    \text{Pr}_\text{f}(\nu \to \nu') = \sum_{\nu''}\text{Pr}_\text{R}(\nu \rightarrow \nu'')\text{Pr}_\text{z}(\nu'' \rightarrow \nu')\,.
\end{equation}
This gives us the explicit transition rules
\begin{equation}
    \text{Pr}_\text{f}(\nu \to \nu-1)= %
    \frac{4^{N}}{(4^{N} - 1)^2}(2^{\nu}-2^{-N})(2^\nu-2^{-\nu})\,,
\end{equation} 
\begin{equation}
    \text{Pr}_\text{f}(\nu \to \nu+1)=
    \frac{4^N-2^{N+\nu}}{4^N-1}\left(1-\frac{2^{N}(2^{\nu+1}-2^{-\nu-1})}{4^{N} - 1}\right)\,
\end{equation}
and
\begin{equation}
\text{Pr}_\text{f}(\nu \to \nu)=1-\text{Pr}_\text{f}(\nu \to \nu-1)-\text{Pr}_\text{f}(\nu \to \nu+1)\,.
\label{eq:pr_f_1_minus}
\end{equation}
These probabilities now define the update rules for the Markov chain for distribution $\rho(\nu)$ via
\begin{equation}\label{eq:markov_magic}
    \rho(\nu) \to \sum_{\nu^\prime} \rho(\nu^\prime) \text{Pr}_\text{f}(\nu^\prime \to \nu)\,.
\end{equation}

Now, we establish an analytical model. For that, let us define the normalized non-stabilized dimension as
\begin{equation}
    w(\nu) \equiv y(\nu)/2^N = 2^{\nu - N} \,.
\end{equation}
In the regime $\{\nu, N\} \gg 1$, the transition probabilities follow
\begin{equation}
    \label{eq:prob_wtow}
     \text{Pr}_\text{f}(w \to w) \approx 3w - 3w^2\,,
\end{equation}
\begin{equation}
    \label{eq:prob_wto2w}
    \text{Pr}_\text{f}(w \to 2w) \approx 1 - 3w + 2w^2 \,,
\end{equation}
and
\begin{equation}
    \label{eq:prob_wtow/2}
    \text{Pr}_\text{f}(w \to w/2) \approx w^2\,.
\end{equation}
At each step of the Markov chain, the mean value $\overline{w}$ changes as
\begin{equation}
\label{eq:magicmeas_update}
    \overline{w} \to 2\overline{w} - 3 \overline{w^2}  + \frac{3}{2} \overline{w^3}\,.
\end{equation}
To derive this, we used the update rule of the Markov chain
\begin{equation}
    \rho(w) \to \sum_{w^\prime} \rho(w^\prime) \text{Pr}_\text{f}(w^\prime \to w)
\end{equation}
for the probability distribution $\rho(w)$, to obtain the corresponding update rule for $\overline{w}$,
\begin{equation}
\begin{aligned}
    \overline{w} \to& \sum_{w,w^\prime} w \rho(w^\prime) \text{Pr}_\text{f}(w^\prime \to w)= \\
    =& \sum_{w} \bigg[w \rho(2w) \text{Pr}_\text{f}(2w \to w) + w \rho(w/2) \text{Pr}_\text{f}(w/2 \to w) \\
    &+ w \rho(w) \text{Pr}_\text{f}(w \to w)\bigg] =\\
    =& \sum_w \rho(w) \left[\frac{w}{2} \text{Pr}_\text{f}(w \to w/2) + 2w \text{Pr}_\text{f}(w \to 2w) \right.
    \\&\left. + w \text{Pr}_\text{f}(w\to w)\right] \\
    =& \frac{1}{2} \overline{w \text{Pr}_\text{f}(w \to w/2)} + 2 \overline{w \text{Pr}_\text{f}(w \to 2w)} + \overline{w \text{Pr}_\text{f}(w\to w)} \\
    =& 2\overline{w} - 3\overline{w^2} + \frac{3}{2} \overline{w^3}\,.
\end{aligned}
\label{eq:markov_magic_approx}
\end{equation}
~\eqref{eq:magicmeas_update} motivates the continuous-time model
\begin{equation}
\label{eq:approx_model_withmagic}
    \frac{d\overline{w}}{dt} \approx \overline{w} - 3 \overline{w}^2  + \frac{3}{2} \overline{w}^3
\end{equation}
analogous to~\eqref{eq:approx_model_nomagic}, assuming that the probability distribution $\rho(w)$ is concentrated close to $\overline{w}$. This model has one stable fixed point at 
$\overline{w} \approx 1 - 1/\sqrt{3} \approx 0.423$. Assuming that $\overline{w} \approx 2^{\overline{\nu}-N}$, this predicts the steady-state value for the mean stabilizer nullity $\overline{\nu} \approx N - 1.24$, i.e., $\overline{\nu}$ is asymptotically maximum up to a small additive shift. Physically, this implies that the injection of magic even for an arbitrarily small $\theta_M$ overwhelms the decay of magic due to the measurement.

We can get further insights on the dynamics by solving the continuous-time model. By inspecting~\eqref{eq:approx_model_withmagic}, we can see that $\overline{w}(t)$ converges to the stable fixed point $\overline{w}_\infty = 1 - 1/\sqrt{3} \approx 0.423$ from above (below) if the initial value $\overline{w}_0$ is above (below) $\overline{w}_\infty$. As we will now show, the convergence timescales in both cases can be drastically different. First, let us consider the case where $\overline{w}_0 > \overline{w}_\infty$, and write 
\begin{equation}
    \overline{w}(T_\downarrow) = w_\infty + \epsilon,
\end{equation}
for some small $\epsilon > 0$. The convergence time, $T_\downarrow$, is given by
\begin{equation}
    T_\downarrow \approx \frac{1}{\sqrt{3} - 1} \ln\left(\frac{\overline{w}_0 - \overline{w}_\infty}{\epsilon}\right),
\end{equation}
which does not explicitly depend on the size of the system $N$. For example, if we initialize in a Haar random state such that $\overline{w}_0 \approx 1$, the convergence time
\begin{equation}
    T_\downarrow \approx \frac{1}{\sqrt{3}-1} \ln\left(\frac{1}{\epsilon}\right)
\label{eq:convergence_time_haar}
\end{equation}
is independent of $N$. On the other hand, if we consider the case where $\overline{w}_0 < \overline{w}_\infty$ and write
\begin{equation}
    \overline{w}(T_\uparrow) = w_\infty - \epsilon,
\end{equation}
again for a small $\epsilon > 0$, the convergence time, $T_\uparrow$, becomes
\begin{equation}
    T_\uparrow \approx \ln\left(\frac{\overline{w}_\infty}{\overline{w}_0}\right) + \frac{1}{\sqrt{3} - 1}\left(\frac{\overline{w}_\infty - \overline{w}_0}{\epsilon}\right).
\end{equation}
For example, if we initialize in a stabilizer state, $\overline{w}_0 = 2^{-N}$, and the convergence time
\begin{equation}
    T_\uparrow \approx N \ln 2 + \frac{1}{\sqrt{3} - 1} \ln\left(\frac{1}{\epsilon}\right)
\label{eq:convergence_time_stab}
\end{equation}
grows linearly with $N$. The difference in timescales between~\eqref{eq:convergence_time_haar} and~\eqref{eq:convergence_time_stab} can be intuitively understood as follows: the steady state value of the stabilizer nullity $\nu$ is extensive in $N$ (with a small constant correction), and each measurement round only changes $\nu$ by at most $1$. Thus, we need $\sim N$ number of measurements to inject extensive magic into the initial stabilizer state, and only a constant number of measurements to remove a constant amount of magic for Haar random states.

Now, we get a more accurate estimate of the steady state value of $\overline{\nu}$ by solving for the steady-state of the Markov chain approximately. To this end, let us denote the steady state probability distribution for $w$ by $\rho_{\infty}(w)$. The steady state satisfies the recurrence relation
\begin{equation}
\begin{aligned}
    \rho_\infty(w) &= \text{Pr}_\text{f}(w \to w)\rho_{\infty}(w) + \text{Pr}_\text{f}(2w \to w)\rho_{\infty}(2w) \\&+ \text{Pr}_\text{f}(w/2 \to w)\rho_{\infty}(w/2)\,.
\end{aligned}
\end{equation}
It is straightforward to see from the Markov chain that $\rho_{\infty}(1) = 0$. Evaluating the recurrence relation for specific values of $w$, we get
\begin{equation}
    \rho_{\infty}(1/4) = \frac{2}{3} \rho_{\infty}(1/2)\,,
\end{equation}
\begin{equation}
    \rho_{\infty}(1/8) = \frac{2}{21} \rho_{\infty}(1/4)\, 
\end{equation}
and
\begin{equation}
    \rho_{\infty}(1/16) = \frac{2}{105} \rho_{\infty}(1/8)\,.
\end{equation}
Now, since $\rho_{\infty}(2/105)$ is very small, we can approximate $\rho_{\infty}(2/105) \approx 0$. Applying the normalization condition $\rho_{\infty}(1/2) + \rho_{\infty}(1/4) + \rho_{\infty}(1/8) \approx 1$, we obtain the steady state probability distribution
\begin{equation}
    \rho_{\infty}(w) \approx \begin{cases}
        0.578, &\quad w = 1/2 \\
        0.385, &\quad w = 1/4 \\
        0.037, &\quad w = 1/8, \\
        0, &\quad \text{else},
    \end{cases}
\end{equation}
rounded to 3 decimal places. From $\rho_{\infty}(w)$, we can estimate
\begin{equation}
    \overline{w} \approx 0.39
\end{equation}
and
\begin{equation}\label{eq:model_distribution_ss}
\nu^\text{ss}_\text{analytic}\equiv\overline{\nu} \approx N - 1.46\,,
\end{equation}
which matches the actual steady-state nullity accurately as we show in the next section.

\subsection{Numerical results}\label{sec:numerical}

\begin{figure}[t!]
    \centering
    \subfigimg[width=0.3\textwidth]{}{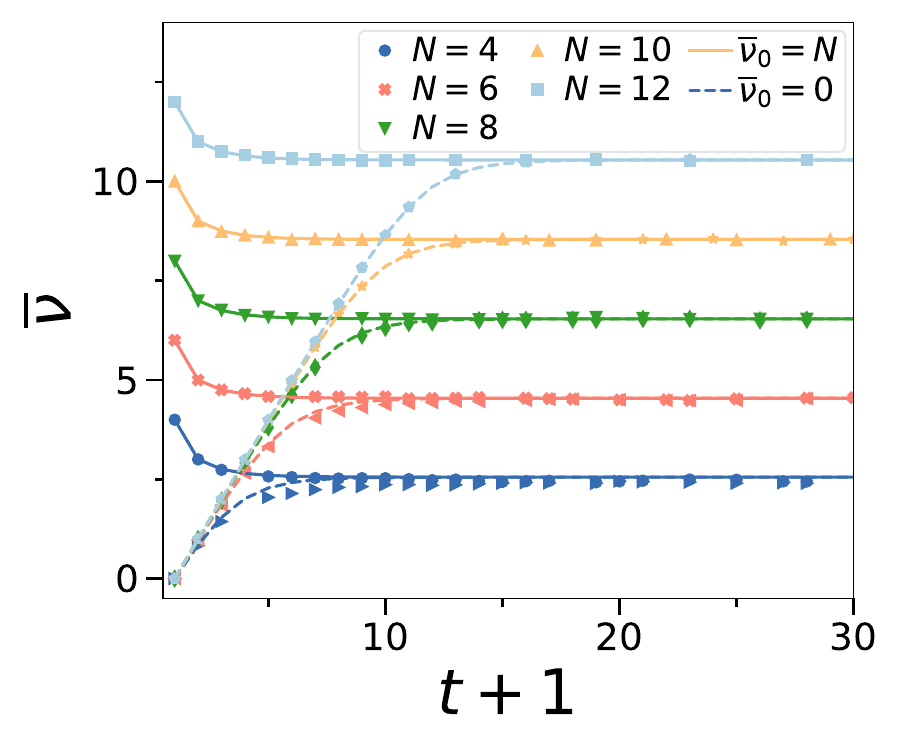}
    \caption{
    Average stabilizer nullity $\overline{\nu}$ against measurement step $t$ for $\theta_M = 0.001$ ($\overline{\nu}$ evolves the same for any $\theta_M>0$). Points show average over $N_r \approx 1000$ trajectories with initial states given by Haar-random states (points starting from $\overline{\nu}=N$) and by stabilizer state $\ket{0}^{\otimes N}$ (starting from $ \overline{\nu}=0$). Solid lines are the numerical simulation of the Markov chain via~\eqref{eq:markov_magic} for initial $\overline{\nu}_0 = N$, while dashed lines are for initial $\overline{\nu}_0 = 0$. 
    } 
    \label{fig:nullity_tau}
\end{figure}

We now compare the analytical predictions with the numerical simulations. 
Considering single trajectories, which we show and analyse in more detail in Appendix~\ref{sec:single_traj}, nullity is bounded by $\nu \leq N-1$ as the measurements reduce $\overline{\nu} \to \overline{\nu} - 1$ whenever $\nu$ approaches the maximum value.
In Fig.~\ref{fig:nullity_tau}, we show how this is reflected in the average $\overline{\nu}$, as it evolves, after a brief transient, toward the steady-state value $\nu^\text{ss}$ which solely depends on $N$ and not on the angle parameter $\theta_M$. 
In this figure, for each fixed size, the points correspond to the value of the average over $N_{\text{r}} \approx 1000$ trajectories of the nullity $\overline{\nu}$ computed numerically for two different classes of states. 
The first class is that of states with maximal initial nullity, $\overline{\nu}_0 = N$, such as $\ket{\psi}_{\text{Haar}}$ states, $\ket{T}^{\otimes N}$ and $\ket{\psi}_{\text{GUE}}$ which all yield the same effective dynamics. The second class starting from $\overline{\nu}_0 = 0$ corresponds to $\ket{0}^{\otimes N}$.

In addition to the exact numerical simulations, we plot in Fig.~\ref{fig:nullity_tau} also numerical simulations of the Markov chain~\eqref{eq:markov_magic}, with the transition probabilities described in~\eqref{eq:pr_r_nu_nu} to~\eqref{eq:pr_f_1_minus}. Here, solid lines for the high-nullity states, and as dashed lines for initial stabilizer states. 

\begin{figure*}[t!]
	\centering	
 	\subfigimg[width=0.3\textwidth]{a}{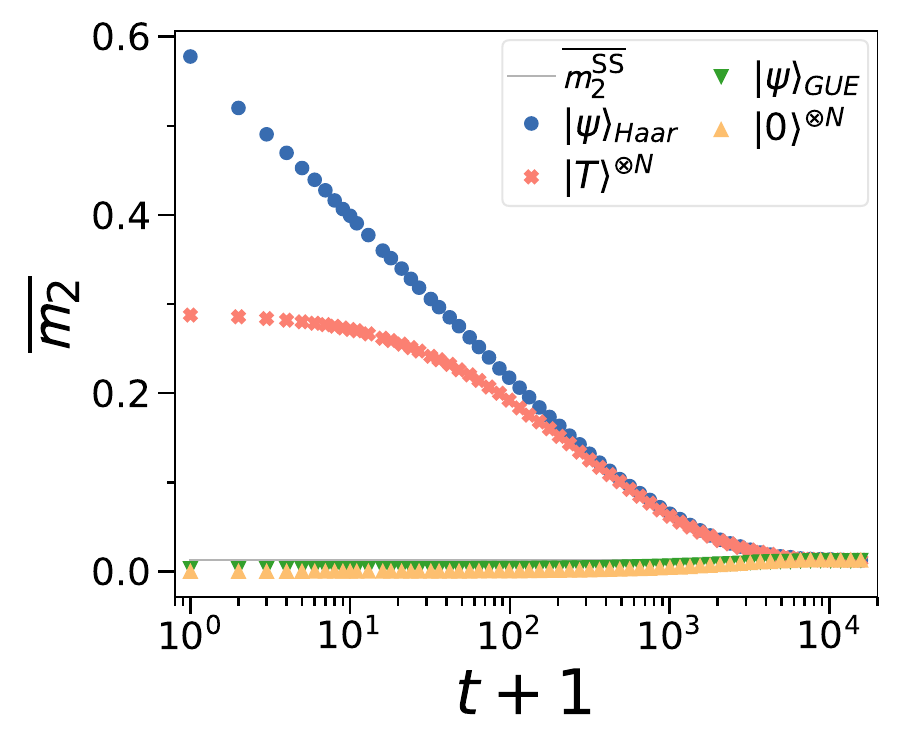}
 	\subfigimg[width=0.3\textwidth]{b}{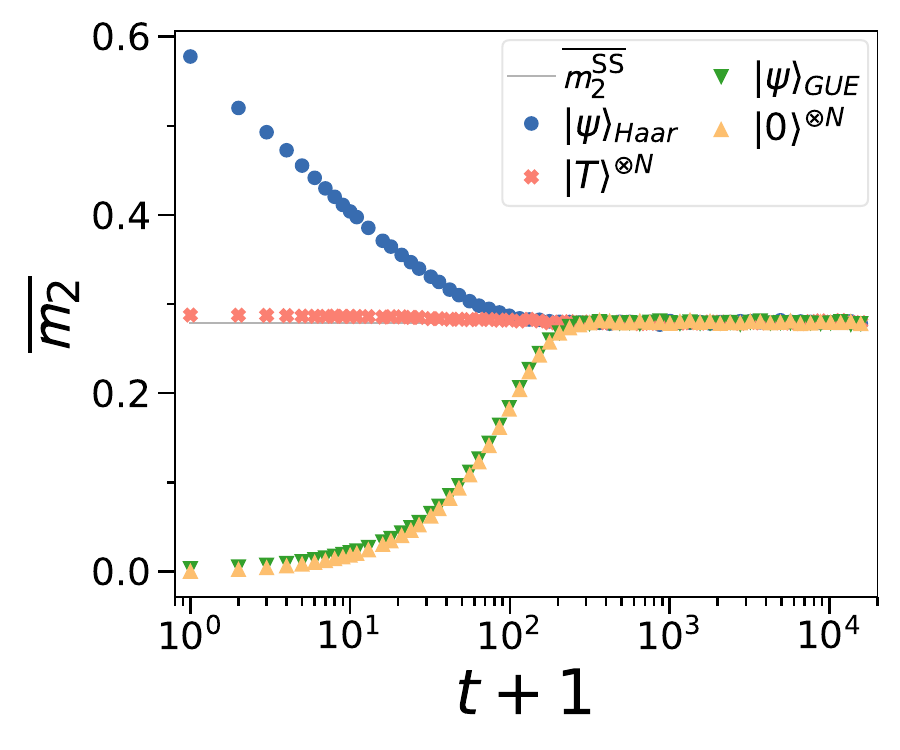}
     \subfigimg[width=0.3\textwidth]{c}{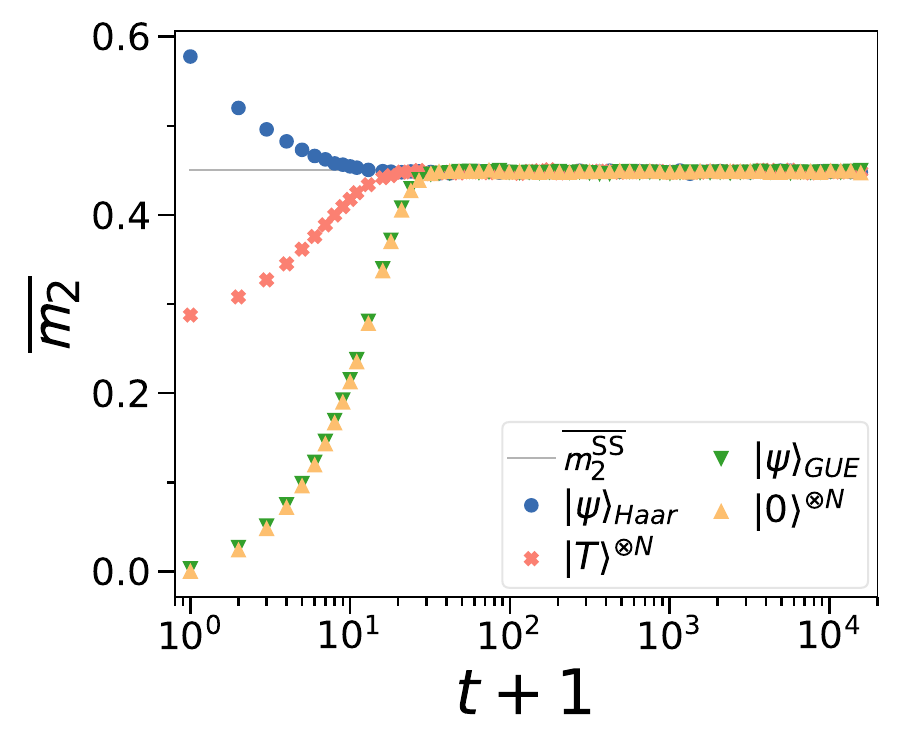}
    \caption{
    \idg{a} Average magic density $\overline{m}_2 = \overline{\mathcal{M}}_2/N$ over $N_\text{r} \approx 1000$ trajectories versus measurement steps for $\theta_M = 0.01$ for $N = 12$ for four different initial states. 
    \idg{b} Average magic density $\overline{m}_2 = \overline{\mathcal{M}}_2/N$ over $N_\text{r} \approx 1000$ trajectories versus measurement steps for $\theta_M = 0.2$ for $N = 12$ for four different initial states. 
    \idg{c} Average magic density $\overline{m}_2 = \overline{\mathcal{M}}_2/N$ over $N_\text{r} \approx 1000$ trajectories versus measurement steps for $\theta_M = 1$ for $N = 12$ for four different initial states. 
	}
    \label{fig:sre2_theta}
\end{figure*}

We numerically confirm that the convergence time to the steady state depends only on the initial nullity: the former states, i.e., initial states with nullity larger than the steady-state value, converge to it in a time $t$ that is independent of $N$, as predicted in~\eqref{eq:convergence_time_haar}. 
In contrast, the latter with lower nullity converge in a time that is linear in $N$, our numerical findings matching the theoretical predictions of~\eqref{eq:convergence_time_stab}. 

Furthermore, the analytical prediction in Sec.~\ref{sec:analytical_magic} gives $\nu^\text{ss}_\text{analytic} \simeq N - 1.46$. Numerically, we find that for all initial states, we converge to the same steady-state with $\nu^\text{ss}_\text{num} \simeq N - 1.46 \pm 0.02$. Thus, we find that our theory and numerics are in excellent agreement.

We now analyze the evolution of the average $\mathcal{M}_2$ for the same set of four different initial states in the presence of measurements in the magical basis ($\theta_M \neq 0$). 
Due to the non-monotonic behavior of SRE along individual trajectories, in the main text we primarily focus on the ensemble-averaged $\overline{\mathcal{M}}_2$ and in particular, we consider the density $\overline{m}_2 = \overline{\mathcal{M}}_2/N$. The behavior of individual trajectories for different $\theta_M$ and initial states is described in detail in Appendix~\ref{sec:single_traj}. 

Unlike stabilizer nullity, which only depends on $\nu_0$ of the initial state, $\overline{m}_2$ strongly depends on the properties of the initial states. This is due to the fact that SREs probe the structure of the Pauli spectrum, which differs for different types of states. However, the asymptotic steady-state is independent of the chosen initial state, and only  depends on the measurement angle parameter $\theta_M$. 
In Figs.~\ref{fig:sre2_theta}a, b and c we show the evolution of the SRE density for three representative values of the parameter $\theta_M = 0.01$, $0.2$ and $1$. 
For the smallest angle $\theta_M = 0.01$ in panel a, the dynamics of both $\ket{\psi}_{\text{Haar}}$ and $\ket{T}^{\otimes N}$ states closely resembles the $\theta_M = 0$ one, while the other two state types have very little increasing SRE until they reach a low steady-state value. Increasing the angle parameter to $\theta_M = 0.2$, in panel b we observe that the number of measurements required to get to the same steady-state decreased, the Haar-random initial states still follow a similar decay profile, while the SRE for $T$-states is nearly constant changing drastically the behavior with respect to the non-magical basis case, and the other two state types behave almost the same increasing monotonically. Finally, at maximal $\theta_M= 1$, all four ensembles rapidly converge to the same steady value within approximately $\sim N$ measurements steps. 
The number of measurements required to reach the steady state strongly depends on the angle parameter $\theta_M$, going from exponentially big for $\theta_M \ll 1$ to linear for $\theta_M \to 1$ in the number of qubits $N$. Despite the difference in the timescales, in the three panels we keep the same range on both axes to better compare the dynamics. 
These different timescales barely depend on the initial state type, although we can still see how SRE of $\ket{T}^{\otimes N}$ states always tend to overlap with the of $\ket{\psi}_{\text{Haar}}$ before reaching the steady state, while the states with initial $\overline{m}_2(0) < \overline{m}_2^{\text{ss}}$
reach it in slightly more measurements than the ones with $\overline{m}_2(0) >\overline{m}_2^{\text{ss}}$.

The continuous growth of the steady state value $\overline{m}_2^{\text{SS}}$ with the angle parameter $\theta_M$ can be better visualized in Fig.~\ref{fig:sre2_theta}b. We find that for `low-magic' basis measurements ($\theta_M \ll1$), the steady-state value increases quadratically with the angle, as shown in the same figure by the dashed fit lines. 
We find that the steady-state is in general not a Haar random state of the form~\eqref{eq:statehaarmeas}, which is evidenced by the fact for any $\theta_M>0$ the nullity is close to $N-1$, while $\mathcal{M}_2$ can be relatively small especially for $\theta_M\ll1$~\cite{gu2023little}. However, the steady state of SREs appears to be close to the corresponding SRE of Haar random states of $N-2$ qubits, as shown in Appendix~\ref{sec:final_value}.

It is important to stress that in this protocol stabilizer nullity $\nu$ is independent of the choice $\theta_M>0$, while SRE strongly depends on $\theta_M$. This suggests that the mapping in~\eqref{eq:avgsre2_avgnullity} is not valid in this case, not even for Haar-random states. Therefore, despite the clarity of the numerical description, further investigations are required to analytically model the behavior of SRE in this protocol. 

\begin{figure}[t!]
	\centering	
 	\subfigimg[width=0.31\textwidth]{}{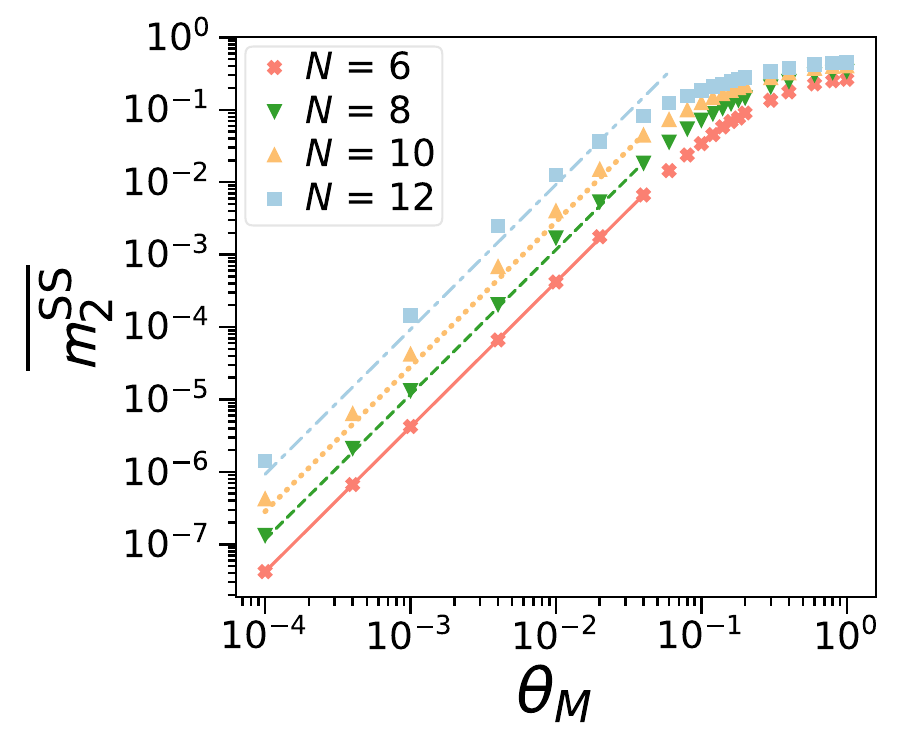}
    \caption{
    Average magic density of the steady state $\overline{m}_2^\text{ss}$ for $t\rightarrow \infty$ as a function of $\theta_M$ for different $N$ shown as dots. For $\theta_\text{M} \ll 1$, we find
    $\overline{m}_2^{\text{ss}} \propto \theta_\text{M}^2$ which is indicated as lines.
	}
    \label{fig:sre2_theta_ss}
\end{figure}

\section{Discussion}\label{sec:discussion}

In this work, we studied how sequences of projective measurements and random Clifford unitaries affect the nonstabilizerness — also known as magic — of quantum states. Starting from Haar-random initial states, we derive an analytical expression for the decay of stabilizer nullity under measurements in the computational basis. While measurements usually destroy magic, we find that random Clifford transformations protect it. 
In fact, we find that erasing all magic requires an exponential number of measurements in the system size, demonstrating the inherent robustness of magic. We note that this matches the time needed for purification~\cite{fidkowski2021dynamical}.
We find that our analytical model for stabilizer nullity $\nu$ as well as for SREs $\mathcal{M}_\alpha$, accurately captures the quantitative behaviour of the dynamics and matches our numerical simulations.

We then extended our analysis to single-qubit measurements in rotated bases, where measurements can also induce magic. Notably, rotated measurements can both generate and suppress magic, which converges to a steady-state for sufficiently long evolution~\cite{haug2024probing}.
We construct an analytical model for the stabilizer nullity $\nu$ in this generalized case, accurately capturing the stabilizer nullity of the ensemble steady state. 
Starting from an initial $\nu$, we observe that the average stabilizer nullity converges to about $N-1$, independent of $\theta_M$ and the initial state. 
Meanwhile, the steady-state value of the SRE $\mathcal{M}_2^\text{ss}$ depends on $\theta_M$, with a quadratic scaling $\mathcal{M}_2^\text{ss} \propto \theta_M^2$ in the small-angle limit. Our findings for measurements in a rotated basis also provide a new lens through which to view the landscape of phases in monitored circuits. Of particular interest is the recently discussed phase of matter characterized by high, volume-law entanglement but low, area-law magic.~\cite{gu2025magicinduced}. Our model provides a tunable protocol to realize and stabilize such a state: by choosing $\theta_M \propto 1/\sqrt{N}$ our circuit naturally flows to a steady state with volume-law entanglement but constant non-zero magic. This provides a concrete and controllable setting to study the properties of this phase.
To support our analytical results, we performed extensive numerical simulations for various initial states, such as Haar-random states, unentangled product states (e.g., $T$-states) and short-time GUE-evolved states. Across all these initial conditions, we observed consistent behavior that confirms the validity of our conclusions and the generality of the observed phenomena.

Our findings reveal that measurements, while typically associated with the destruction of quantum resources, can also act as a source of magic under suitable conditions. This highlights new avenues for manipulating nonstabilizerness in hybrid quantum protocols and raises intriguing questions about the interplay between measurement, randomness, and quantum computational resources.

\begin{acknowledgments}
We thank Lorenzo Piroli, Myungshik Kim, Procolo Lucignano and Lorenzo Leone for insightful discussions.
A.S. gratefully acknowledges the hospitality of the Technology Innovation Institute (TII) in Abu Dhabi, where this work was conceived and partially carried out.
M.C. was supported by the PNRR MUR project PE0000023-NQSTI, and the PRIN 2022 (2022R35ZBF) - PE2 - ``ManyQLowD''.

\end{acknowledgments}

\bibliography{bibliography}

\onecolumngrid

\let\addcontentsline\oldaddcontentsline
\appendix
\onecolumngrid
\newpage 

\setcounter{secnumdepth}{2}
\renewcommand{\thesection}{\Alph{section}}
\renewcommand{\thesubsection}{\arabic{subsection}}
\renewcommand*{\theHsection}{\thesection}

\clearpage
\begin{center}

\textbf{\large Appendix}
\end{center}
\makeatletter

We provide proofs and additional details supporting the claims in the main text.

\makeatletter
\@starttoc{toc}
\makeatother

\section{Single trajectories in magical measurements}\label{sec:single_traj}
We now study individual trajectories for our circuit where we consider measurements in rotated basis, i.e. $\theta_M>0$.
In Fig.~\ref{fig:circuit}b and c we show a single trajectory and the average value of the stabilizer nullity $\nu$ and von Neumann entanglement entropy $\mathcal{S}$ for initial Haar-random states for measurements in the computational basis. 
In this section, we show how these quantities change in single trajectories and on average with this measurement protocol in the magic basis and how the angle parameter $\theta_M$ affects the outcome of the procedure. 

\begin{figure}[htbp]
    \centering
	\centering	
 	\subfigimg[width=0.3\textwidth]{a}{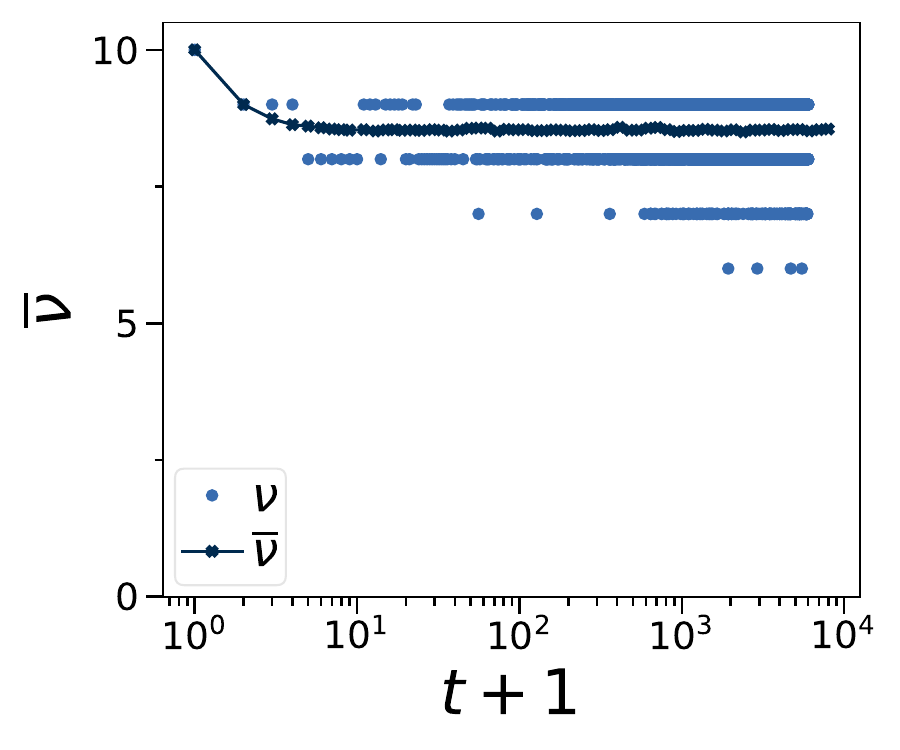}
    \subfigimg[width=0.3\textwidth]{b}{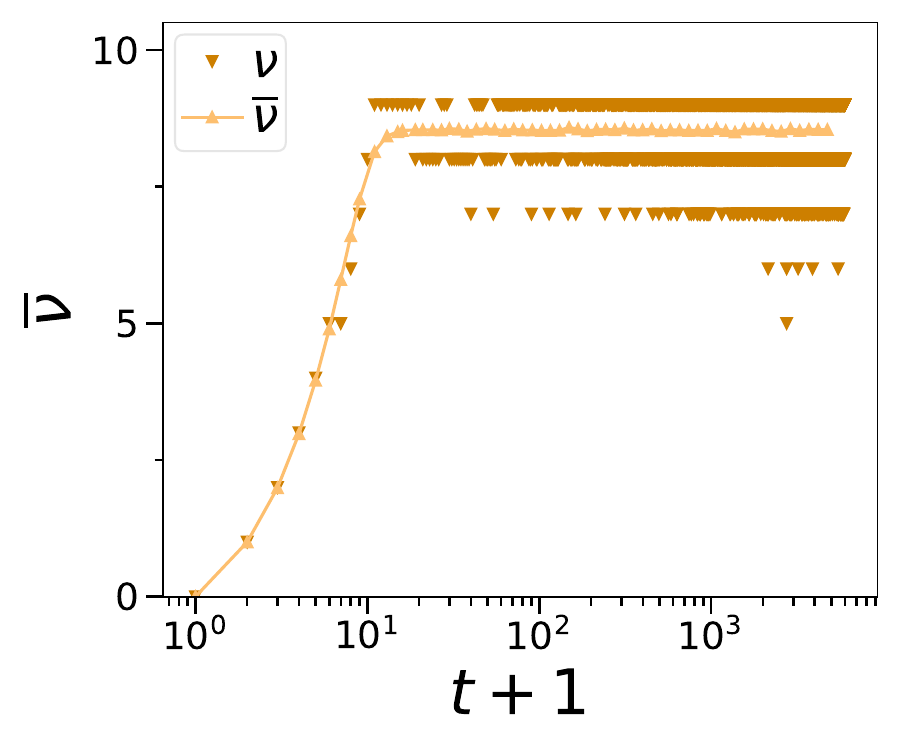}
    \caption{
    \idg{a} Average $\overline{\nu}$ and a single random trajectory $\nu$ as a function of measurements $t+1$ for $N = 10$ and $\theta_M = 0.0001$ starting from Haar-random states.
    \idg{b} Average $\overline{\nu}$ and a single random trajectory $\nu$ as a function of measurements $t+1$ for $N = 10$ and $\theta_M = 0.0001$ starting from $\ket{0}^{\otimes N}$ states.
    }
    \label{fig:nullity_thetam}
\end{figure}

When measuring in the rotated bases, since the nullity evolution does not depend on the value of $\theta_M$, we can see that nullity will oscillate regardless of the angle around the asymptotic value $\nu^{\text{ss}}_{\text{analytic}}$ creating plateaus for $\nu = N-1\,, N-2$, (and with low probability $N-3$ and lower), which do not change with the angle. 
Two examples, one for initial Haar-random states and one for $\ket{0}^{\otimes N}$ states are shown in Fig.~\ref{fig:nullity_thetam}a,b. The dynamics does not depend on the value of $\theta_M$, and we show $\theta_M=0.0001$ to highlight the non-perturbative behavior of the stabilizer nullity.

\begin{figure}[t!]
	\centering	
 	\subfigimg[width=0.25\textwidth]{a}{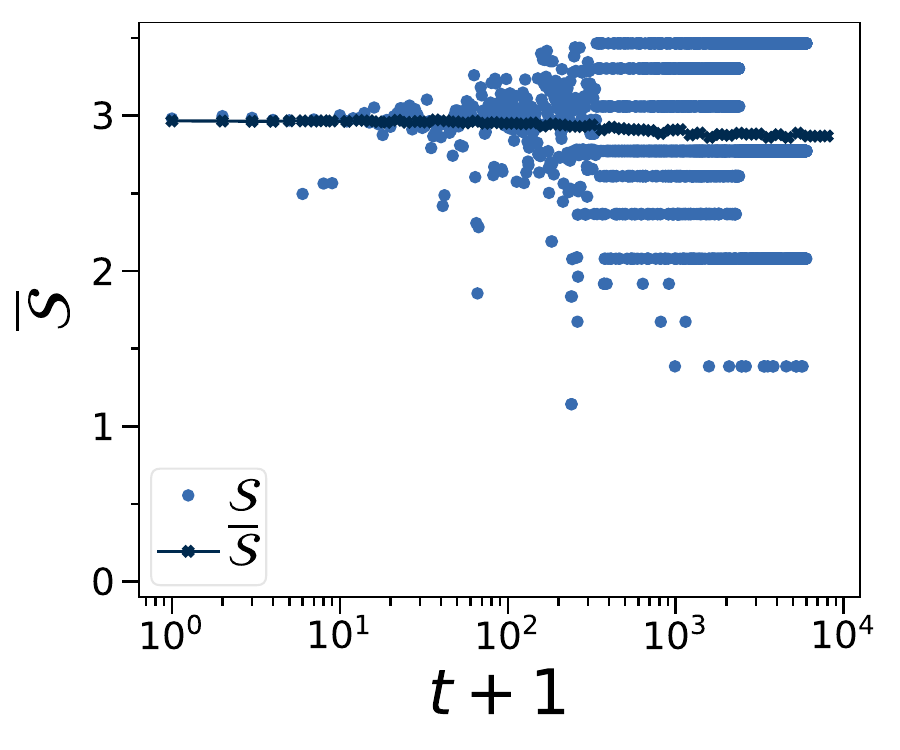}
    \subfigimg[width=0.25\textwidth]{b}{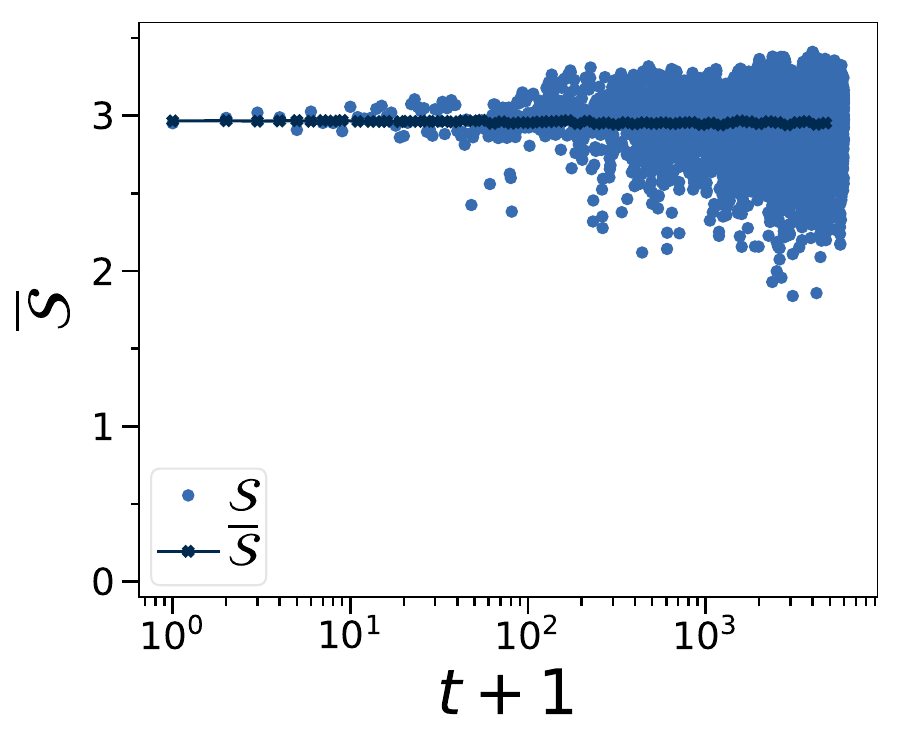}
 	\subfigimg[width=0.25\textwidth]{c}{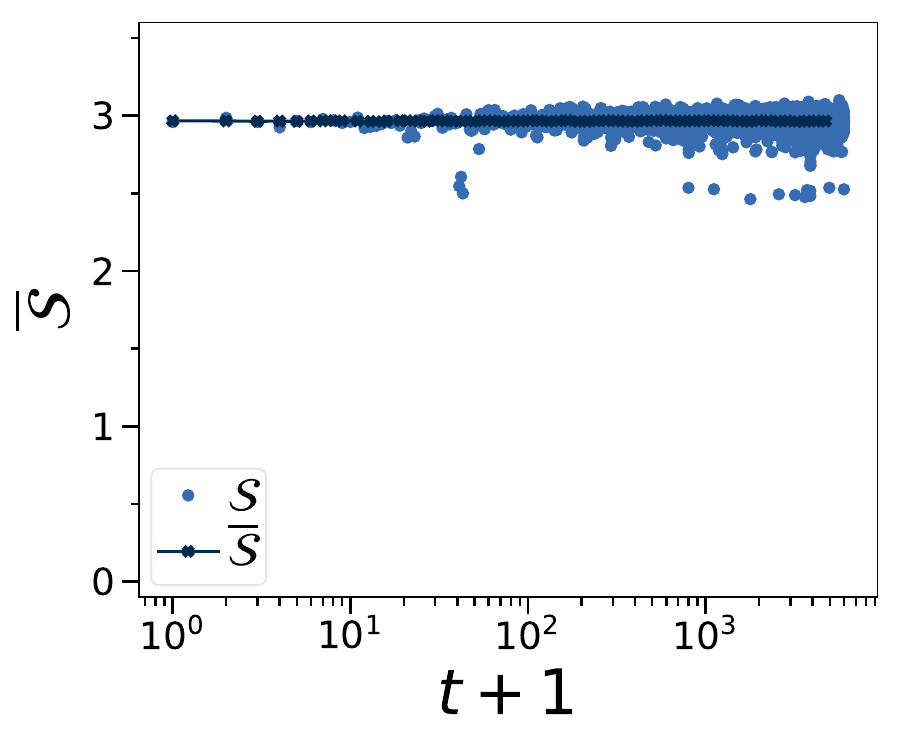} \\    
 	\subfigimg[width=0.25\textwidth]{d}{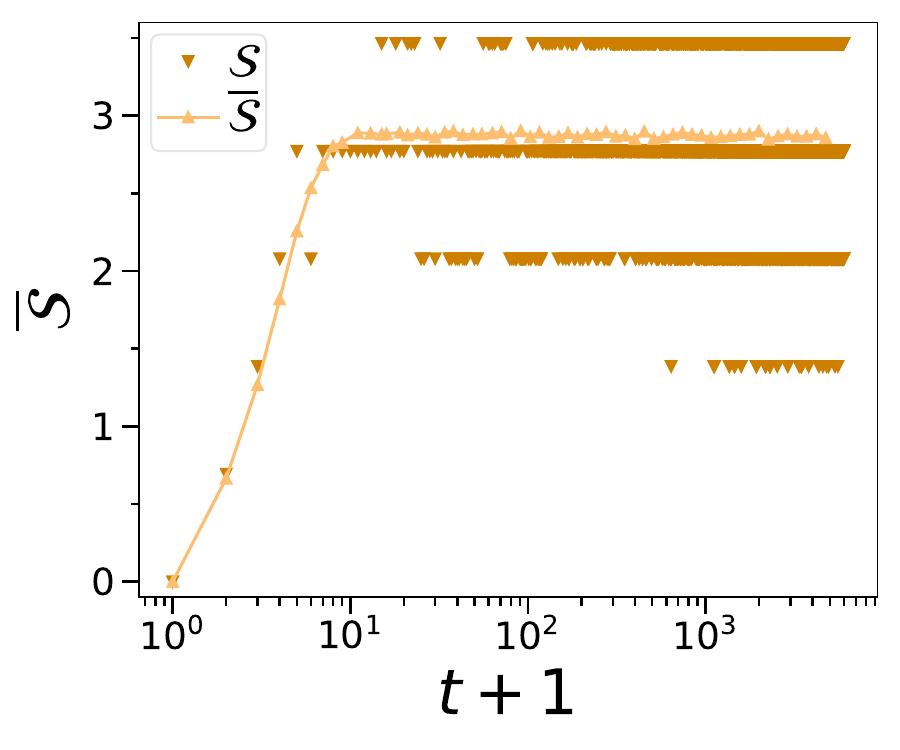}
 	\subfigimg[width=0.25\textwidth]{e}{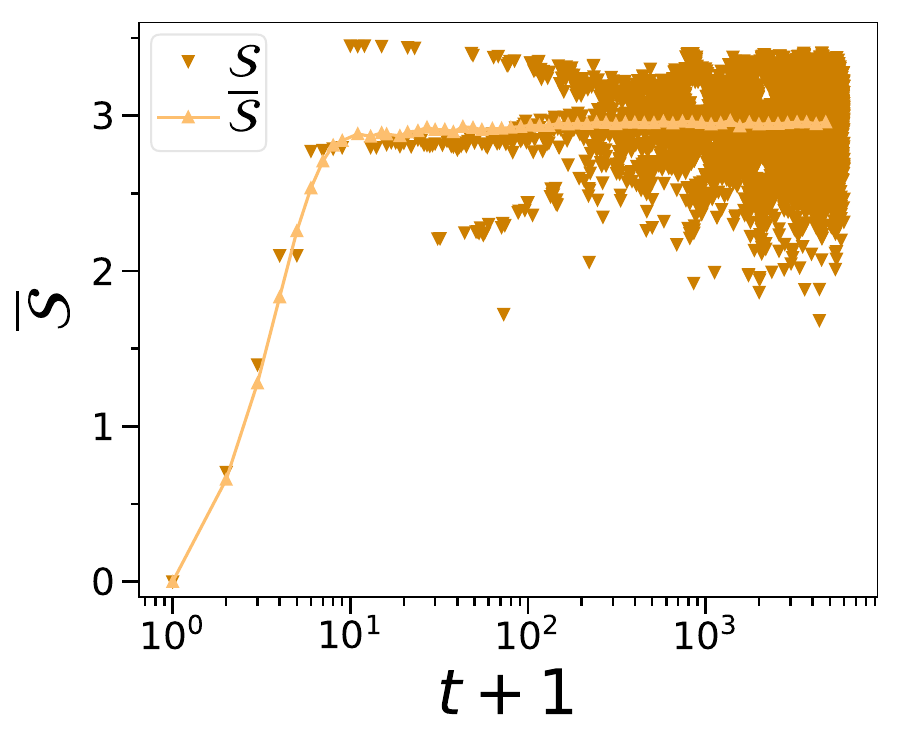}
 	\subfigimg[width=0.25\textwidth]{f}{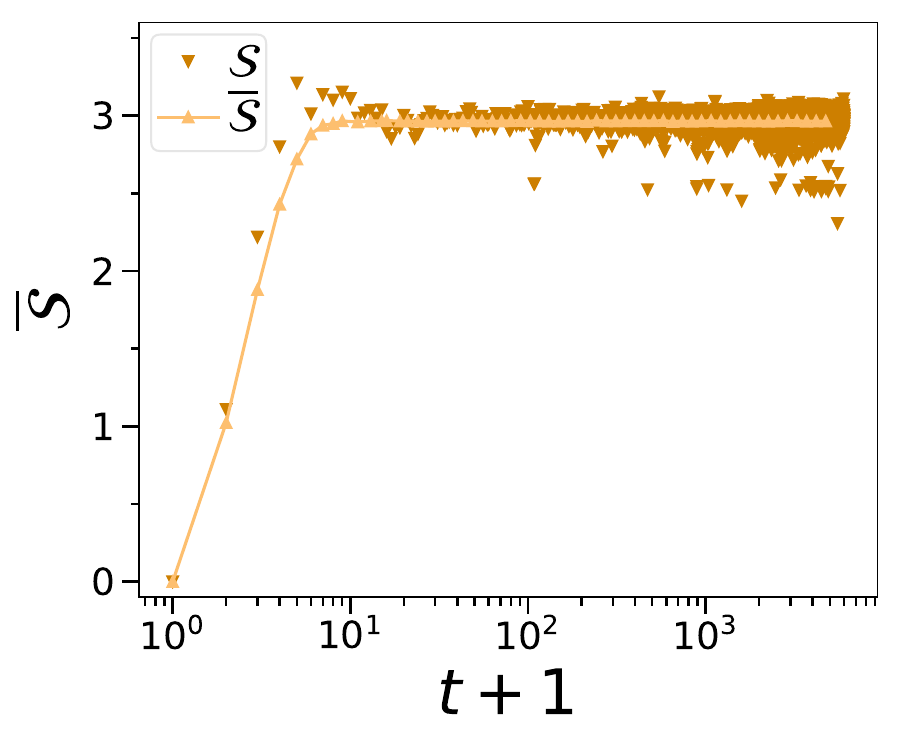} \\
    \caption{
    Von Neumann entanglement entropy $\mathcal{S}$ for $N = 10$ qubits. The first row (a,b,c) contains Haar-random initial states, the second row (d,e,f) all $\ket{0}^{\otimes N}$. The first column (a,d) contains as dotted a random trajectory for $\theta_M = 0.0001$ and as a line the average over $N_{\text{r}} \sim 1000$ different evolutions: the plateau structure is still visible. The second column (b,e) contains the same information but for $\theta = 0.1$. In the last column (c,f) we show single trajectories and the average for the same three states for $\theta_M = 1$: the plateau structure is lost and the entanglement is concentrated around its average value. 
	}
    \label{fig:vnentropy_theta}
\end{figure}

\begin{figure}[htbp]
	\centering	
 	\subfigimg[width=0.3\textwidth]{a}{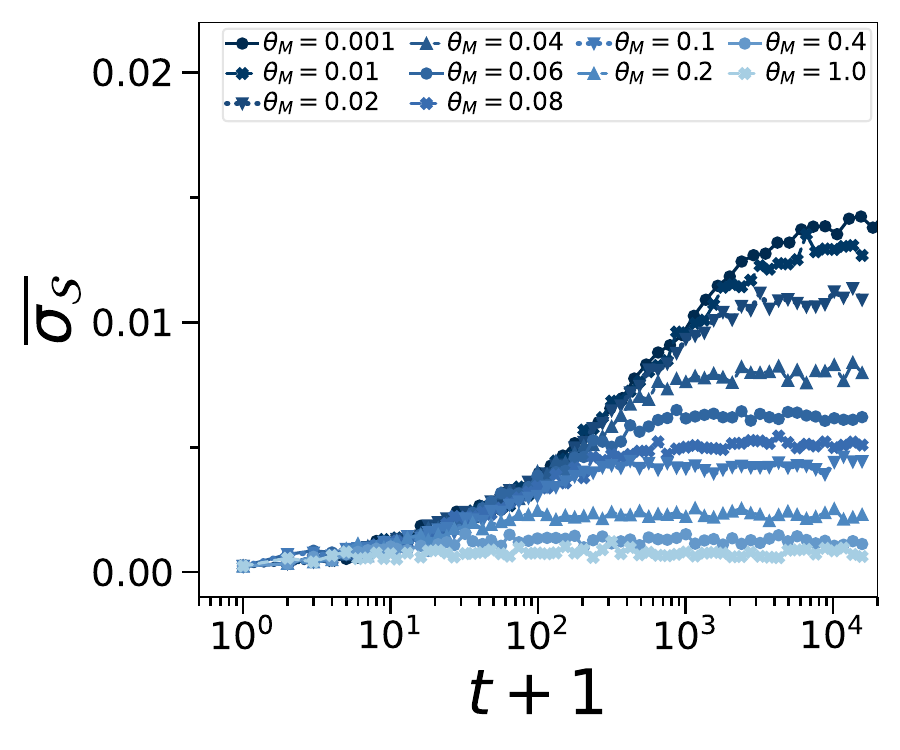}
 	\subfigimg[width=0.3\textwidth]{b}{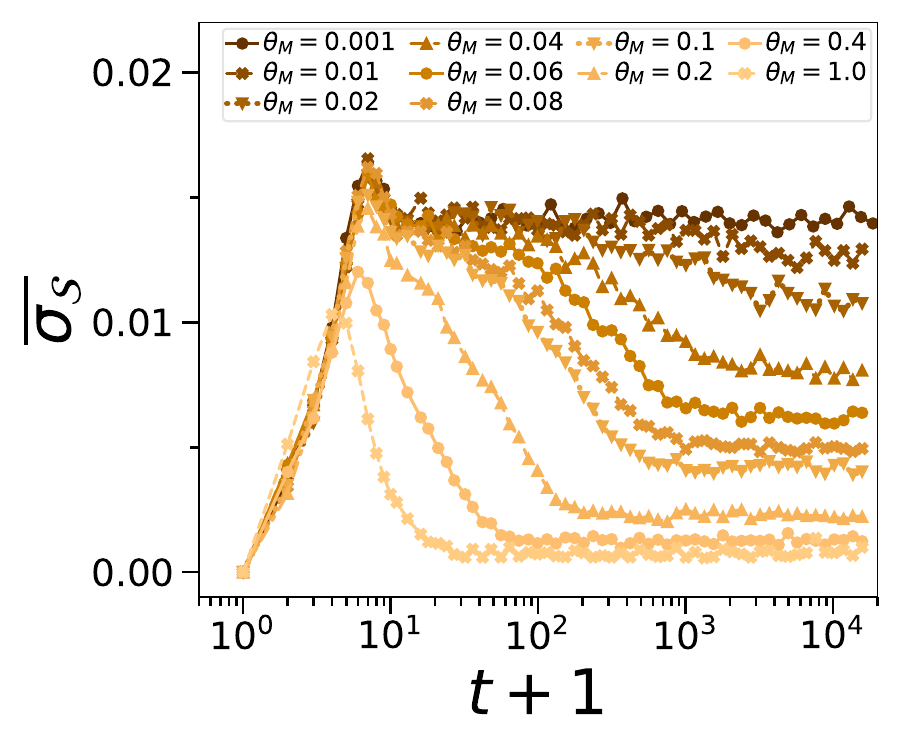}
    \subfigimg[width=0.3\textwidth]{c}{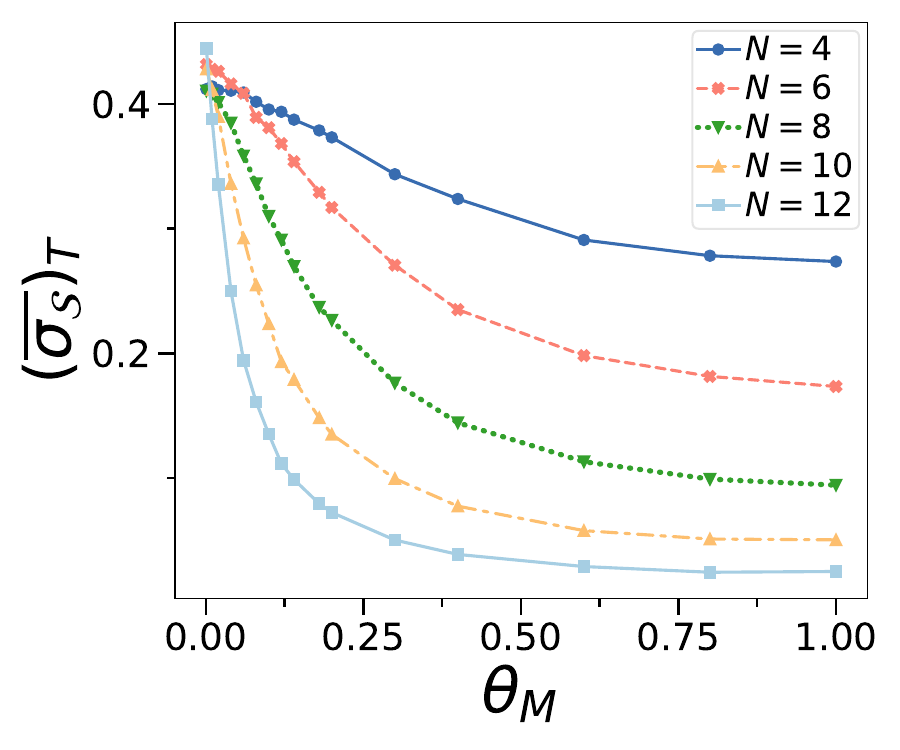}
    \caption{ Standard deviation $\sigma_{\mathcal{S}}$ of von Neumann entanglement entropy.
    \idg{a} $\overline{\sigma_{\mathcal{S}}} = \sigma_{\mathcal{S}}/\sqrt{N_\text{r}}$ vs $t+1$ for initial Haar-random states for $N = 12$ qubits, $N_r = 1000$, for different angle parameters fixed.
    \idg{b} $\overline{\sigma_{\mathcal{S}}} =\sigma_{\mathcal{S}}/\sqrt{N_\text{r}}$ vs $t+1$ for initial $\ket{0}^{\otimes N}$ states for $N = 12$ qubits, $N_r = 1000$.).
    \idg{c} Standard deviation $\overline{\sigma_{\mathcal{S}}}$ averaged over time $T$ as shown in~\eqref{eq:avgstdEE}. We plot against angle parameter $\theta_M$ for different qubit numbers $N$ for Haar-random initial states. 
    }
    \label{fig:std_vne}
\end{figure}

We already showed that at the end of the procedure described in Sec.~\ref{sec:analytical_model}, the resulting state is a random stabilizer state, therefore its entanglement entropy can only assume discrete values. In each panel of Fig.~\ref{fig:vnentropy_theta} it showed both a single representative trajectory and the average over $N_{\text{r}} \approx 1000$ different trajectories of entanglement entropy. In particular, the blue plots in the first row represent the said behavior for Haar-random states, while the yellow plots in the second row represent what happens for $\ket{0}^{\otimes N}$ states. In the plots (a,d), we show what happens for $\theta_M = 0.0001$: as one can see, starting from both Haar-random states and from the computational basis state, the typical plateau structure we see when the state approaches a random stabilizer. Moving to (b,e), for $\theta_M = 0.1$ the plateau structure begins to fade and the entanglement entropy at the end of the dynamics assumes more values continuously spread around its average value, computed over $N_{\text{r}}$ different realizations. 
For the maximal value $\theta_M = 1$, the plateau structure is totally lost, and the entanglement entropy is concentrated very close to the average value. This can be visually seen by looking at the standard deviations of the mean $\overline{\sigma_{\mathcal{S}}} = \sigma_{\mathcal{S}}/\sqrt{N_{\text{r}}}$ in Fig.~\ref{fig:std_vne}a for Haar-random states and in Fig.~\ref{fig:std_vne}b for initial product states $\ket{0}^{\otimes N}$, for $N = 12$ qubits. 
We see that the steady-state value of the standard deviations are independent of the initial-state, which we show in Fig.~\ref{fig:std_vne}c. In all cases the $\overline{\sigma_\mathcal{S}}$ decreases with increasing angle parameter, and while fixing the angle parameter it decreases with increasing system size $N$. The average over time is
\begin{equation}\label{eq:avgstdEE}
    (\overline{\sigma_{\mathcal{S}}})_T = 1/T \int_{t_\text{min}}^{t_\text{max}} \overline{\sigma_\mathcal{S}} dt\,,
\end{equation}
where the integration time $T = t_\text{max} - t_\text{min}$, $t_{\text{max}}$ is the end of the evolution, and $t_\text{min} = N$. 

\begin{figure}[htbp]
	\centering	
 	\subfigimg[width=0.25\textwidth]{a}{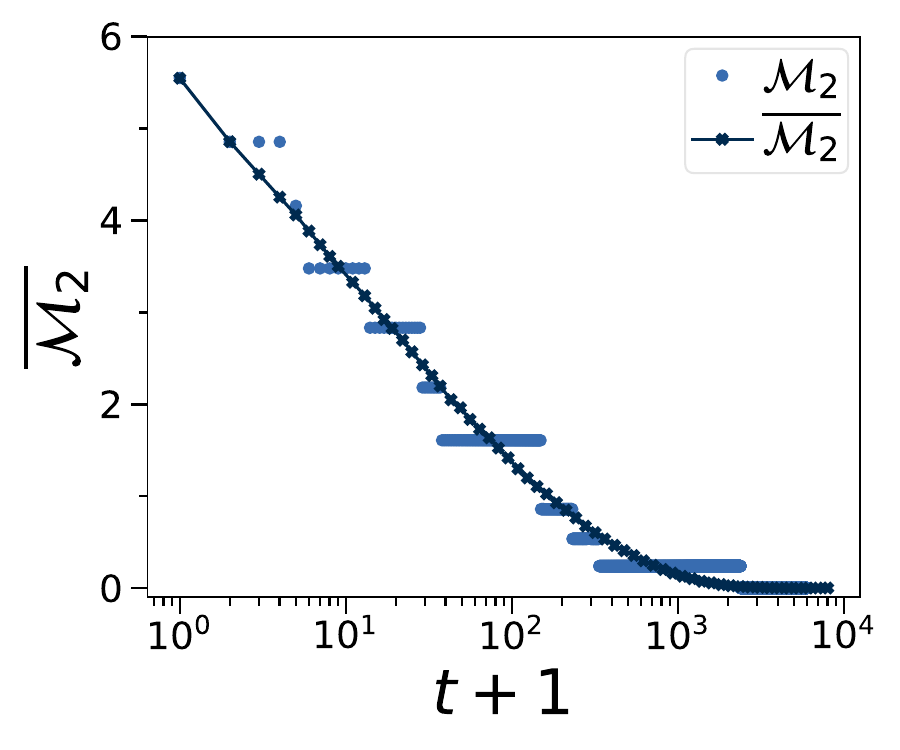}
 	\subfigimg[width=0.25\textwidth]{b}{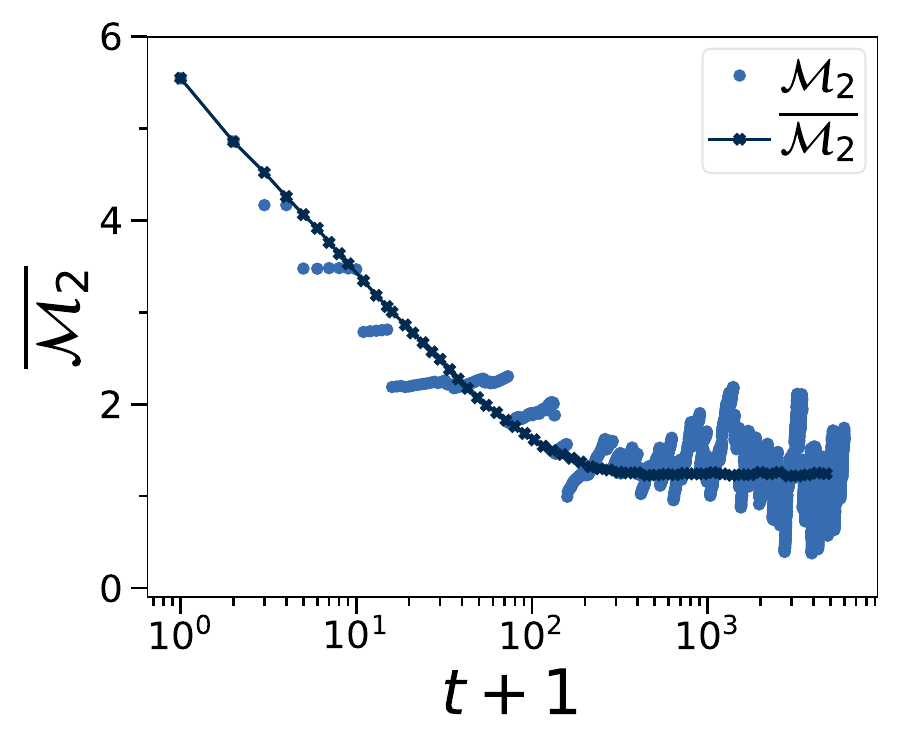}
    \subfigimg[width=0.25\textwidth]{c}{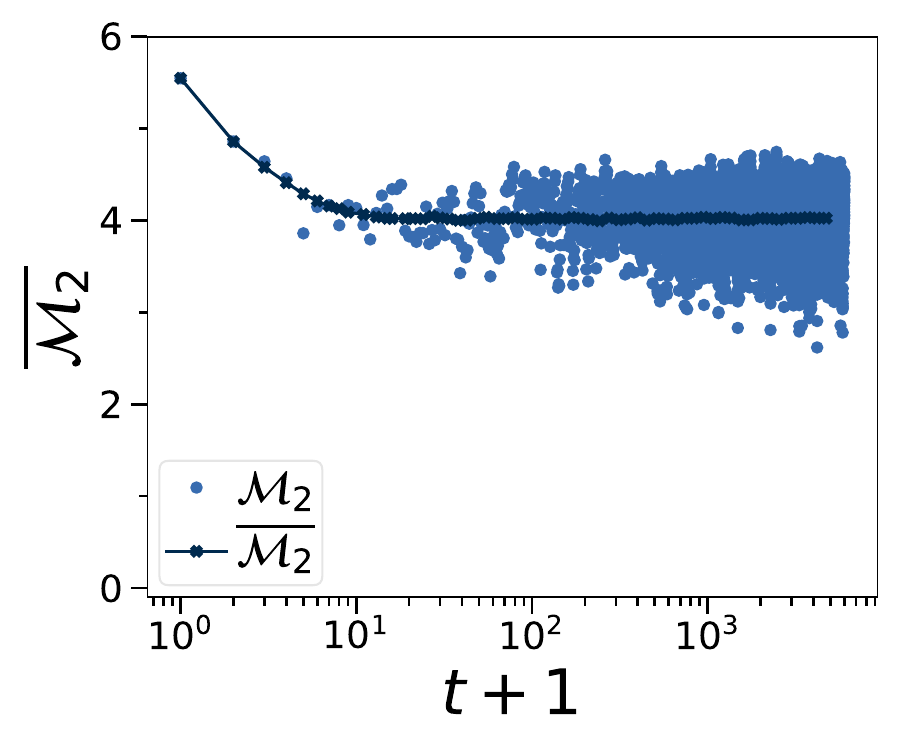}
 	\subfigimg[width=0.25\textwidth]{d}{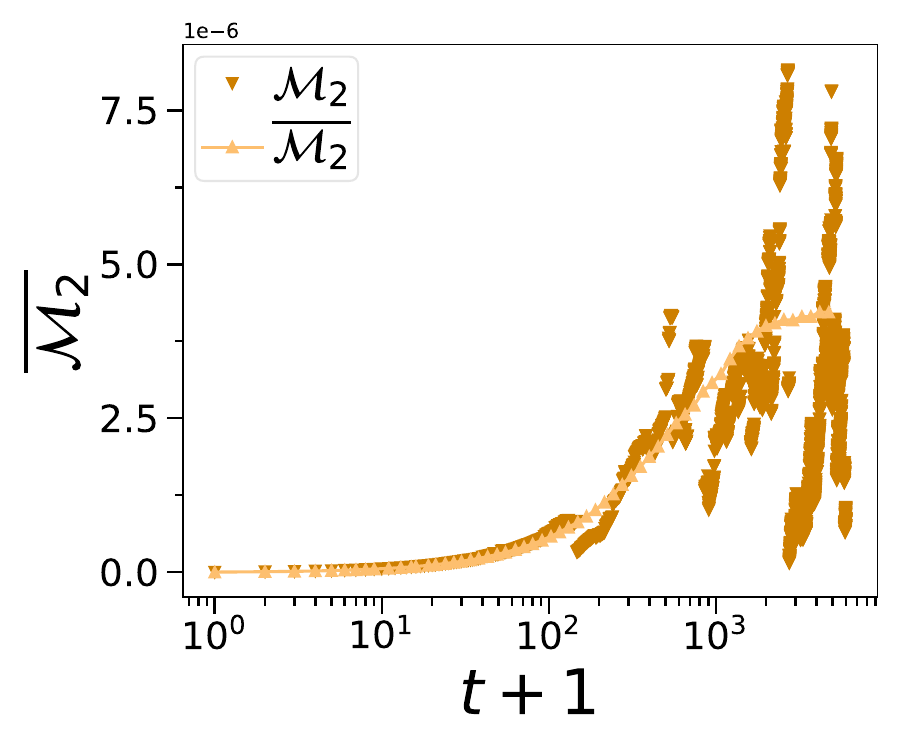}
 	\subfigimg[width=0.25\textwidth]{e}{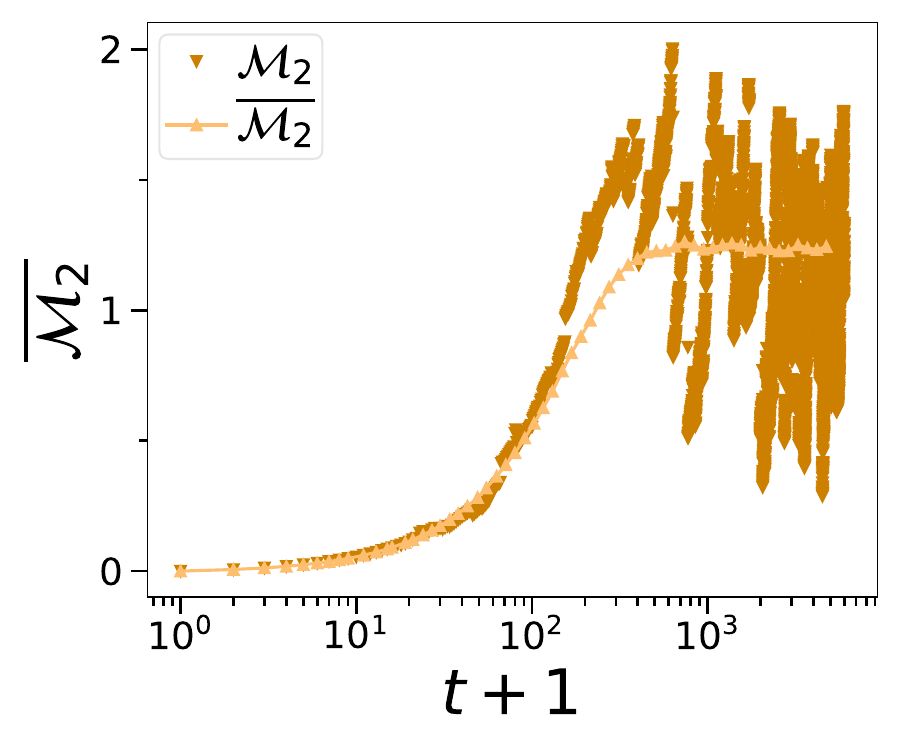}
    \subfigimg[width=0.25\textwidth]{f}{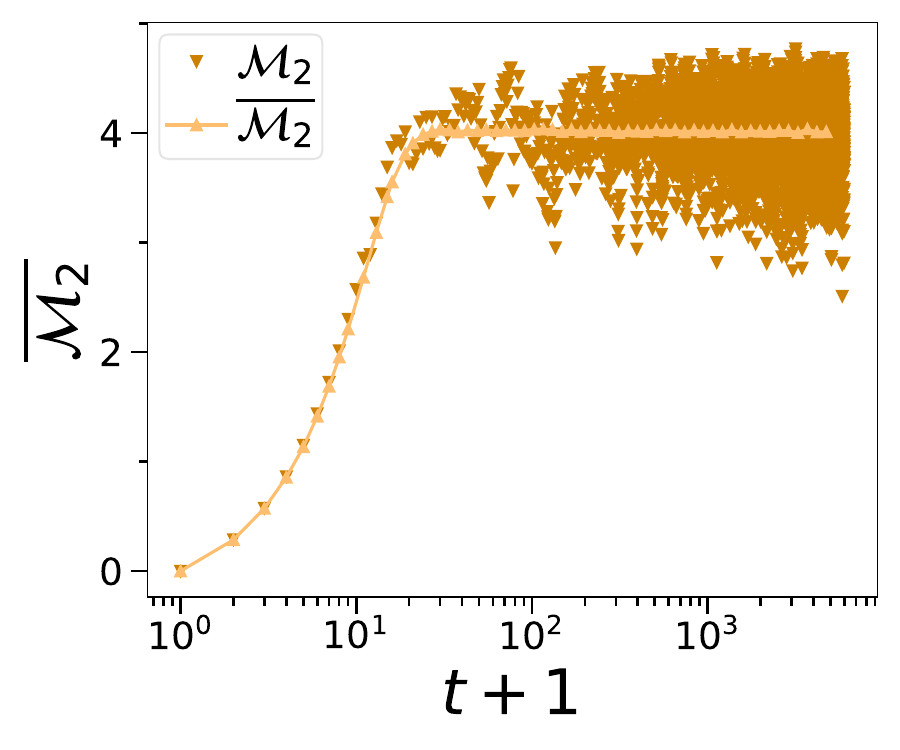}
    \caption{SRE trajectories for $N = 10$ qubits. The first row (a,b,c) contains Haar-random initial states, the second row (d,e,f) all $\ket{0}^{\otimes N}$. The first column (a,d) contains as dotted a random trajectory for $\theta_M = 0.0001$ and as a line the average over $N_{\text{r}} \sim 1000$ different evolutions.}
    \label{fig:sre2_theta_single_traj}
\end{figure}

Next, we study trajectories of our circuit for the SRE. The behavior of the SRE is not very different from the nullity, except that the difference between two plateaus is not a fixed integer as shown in Fig.~\ref{fig:sre2_theta_0}(a). 
Note that the measurement protocol with $\theta_M > 0$ changes the single trajectories and therefore the average behavior. Also, it is worth noting that for every $\theta_M > 0$ the nullity changes on timescales which depend linearly on $N$, while for the SRE we still need exponential timescales for $\theta_M < 1$.
In Fig.~\ref{fig:sre2_theta}(a,b,c) we show how the SRE converges to the steady-state value for the different initial states, therefore in Fig.~\ref{fig:sre2_theta_single_traj} we show how this average dynamics compares to single evolutions.
For Haar-random states in Fig.~\ref{fig:sre2_theta_single_traj}(a,b,c) for increasing angle parameters, it is evident that the plateau structure the dynamics is conserved up to the moment the SRE reaches its angle-dependent steady-value. For $\theta_M = 0.1$ in Fig.~\ref{fig:sre2_theta_single_traj}(b) we can clearly see that the moment the average approaches its steady value coincides with the moment the plateau structure is loose. For $\theta_M = 1$ in Fig.~\ref{fig:sre2_theta_single_traj}(c), the SRE takes $\sim N$ measurements to reach $\mathcal{M}_2^{\text{ss}}$, which means there is no time to create plateaus. 

We repeat the same analysis for $\ket{0}^{\otimes N}$ in Fig.~\ref{fig:sre2_theta_single_traj}(d,e,f), where the initial state has no entanglement or magic. In this case, as $\mathcal{M}_2^{\text{ss}}(\theta_M \ll 1)\ll 1$. For $\theta_M \ll 1$, we can see that measurement in the computational basis increases the SRE by a very small amount, which could be due to the structure of the entanglement entropy for such state as shown in Fig.~\ref{fig:vnentropy_theta}(d): such state is still very close to a stabilizer. Anyway, for $\theta_M < 1$ we can see that still the majority of measurement steps increases magic which then rapidly drops and grows again. 
For $\theta_M = 1$, we see that the SRE reaches its steady-value in $\sim N$ measurement and then oscillates around it. 

\section{Increase in SRE after measurement}\label{sec:magic_increase}

Under Clifford measurements, a (strong) magic monotone is expected to not increase on average. However, when looking at individual trajectories, magic can actually increase, as long as it is compensated by a more significant decrease by another trajectory. As SREs are not strong monotones, such increases can happen even when averaging over trajectories~\cite{haug2023stabilizer}.

\begin{figure}[t!]
	\centering	
 	\subfigimg[width=0.31\textwidth]{a}{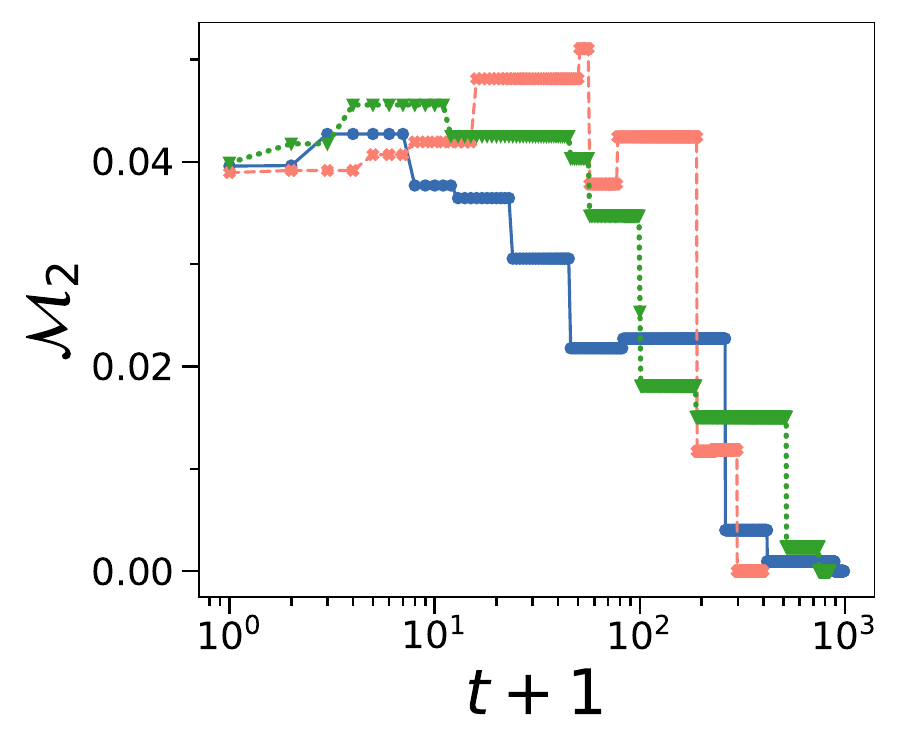}
 	\subfigimg[width=0.31\textwidth]{b}{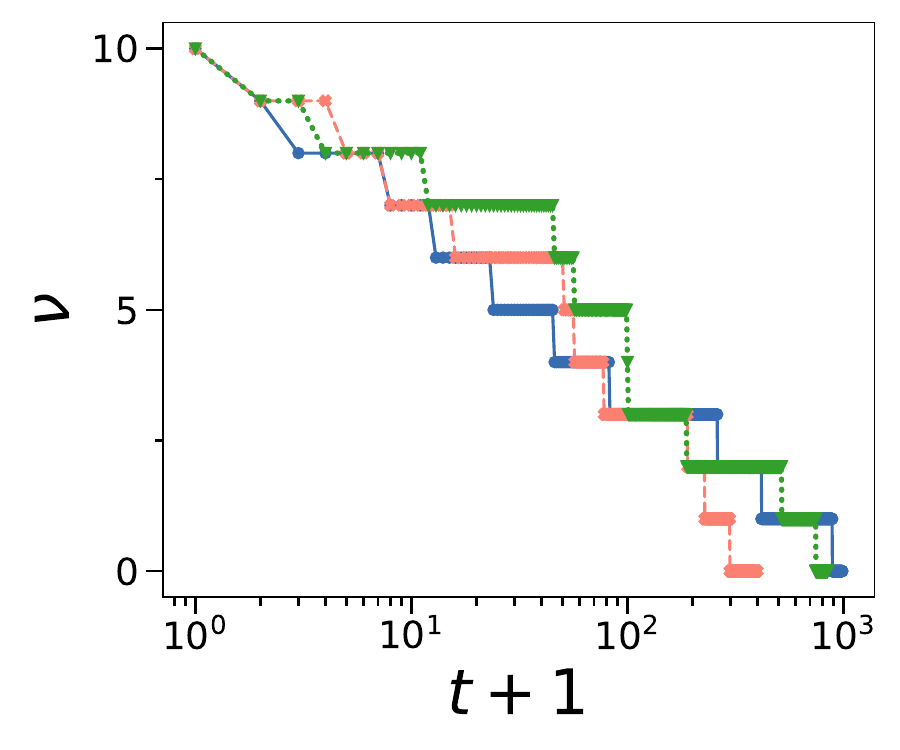}
    \caption{
    \idg{a} $\mathcal{M}_2$ versus measurement $t+1$ computed for three different trajectories starting from $\ket{\psi}_{\text{GUE}}$,
    for $N = 10$ and $\theta_M=0$.
    \idg{b} $\nu$ versus measurement $t+1$ computed for the same trajectories as in panel (a).
	}
    \label{fig:M2_vs_nullity}
\end{figure}

We observe significant increases in SREs in rare instances of trajectories specifically for  $\ket{\psi}_{\text{GUE}}$ states.
As example, we plot the SRE $\mathcal{M}_2$ and nullity $\nu$ for three different random trajectories starting from three different states in Fig.~\ref{fig:M2_vs_nullity}(a,b), in which the same color curves correspond to the same initial state and same evolution. 
As can be seen, in Fig.~\ref{fig:M2_vs_nullity}(a), while all states at the beginning have more or less the same $\mathcal{M}_2 \approx 0.04$, throughout the evolution of $t \sim 10^3$ measurements the SRE increases after a measurement with very low probability ($\approx 1/10^3$).
For the same trajectories, in the correspondence of the change-inducing measurement, $\nu$ decreases. 
Note that such outlies are averaged out over many trajectories, and we find a monotone decrease in SRE on average under Clifford measurements.

Magic monotones must not increase under Clifford channels, while strong monotones must not increase when average over measurement outcomes. 
However, individual trajectories are allowed to increase in magic for magic monotones. For the SRE, we observe such increase in magic rarely, where we can observe quite dramatic spikes in SRE, with a factor $\sim 100$ times increase in SRE after measurement. Notably, we find that such spikes in magic are initial state dependent, where we observe them most notably for the GUE initial state. 
In contrast, we do not see any such spikes for Haar random states, while for $\ket{T}^{\otimes N}$ states we find very small increases in SRE with very small probability. 

\section{Comparison numerical and analytic results}
\label{sec:comparison}

We show a more complete comparison between the analytical predictions in Sec.~\ref{sec:analytical_model} and the numerical data. We already showed that the probability of decay in~\eqref{eq:nullitydecay} correctly predicts the decay of stabilizer nullity as shown in Fig.~\ref{fig:nullity}(b). Now we want to explore the correctness of the predictions of the differential equation in~\eqref{eq:approx_model_nomagic} derived from the decay probabilities. 

\begin{figure}[htbp]
    \centering
 	\subfigimg[width=0.3\textwidth]{}{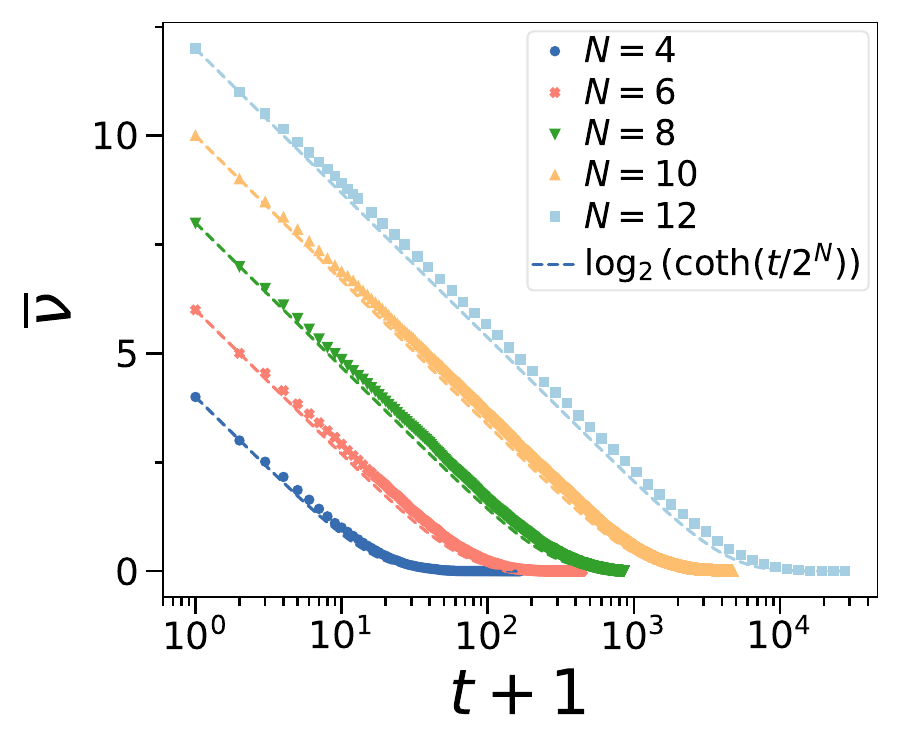}
    \caption{Comparison of the average stabilizer nullity $\overline{\nu}$ for different sizes with the approximate differential equation solution in~\eqref{eq:approx_nullity_nomagic} for different sizes $N$ as a function of measurement steps $t+1$.}
    \label{fig:compare_cotan}
\end{figure}

In Fig.~\ref{fig:compare_cotan} we directly compare the approximation in~\eqref{eq:approx_nullity_nomagic} for large $N$ and $t \gg 2^N$ with the numerical simulation, and this approximate solution already is in good agreement with the numerical data.

\begin{figure*}[t!]
	\centering	
    \subfigimg[width=0.3\textwidth]{a}{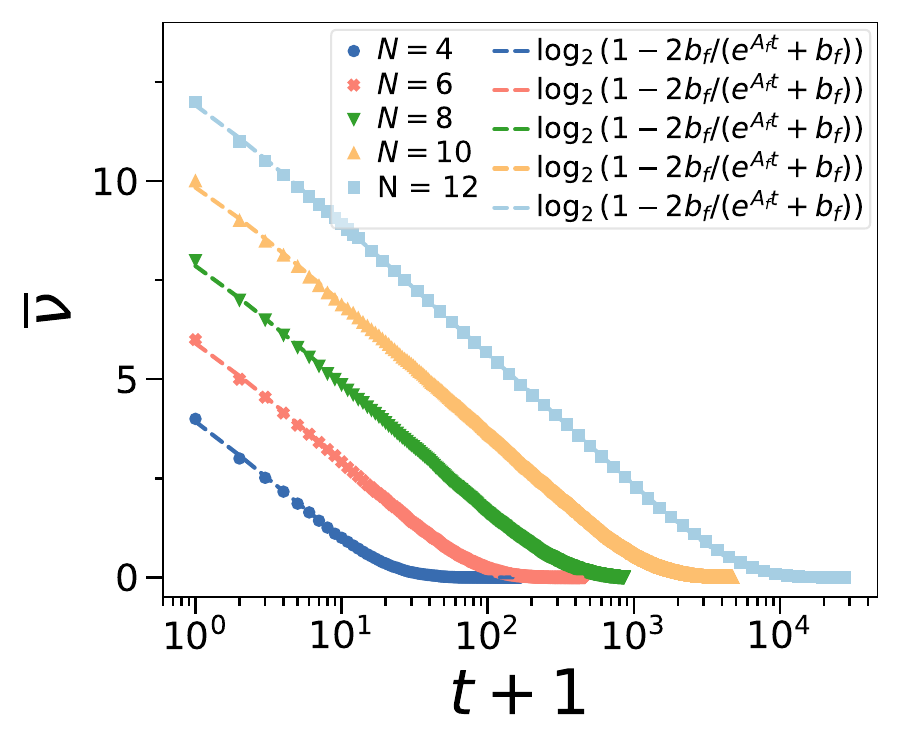}
    \subfigimg[width=0.3\textwidth]{b}{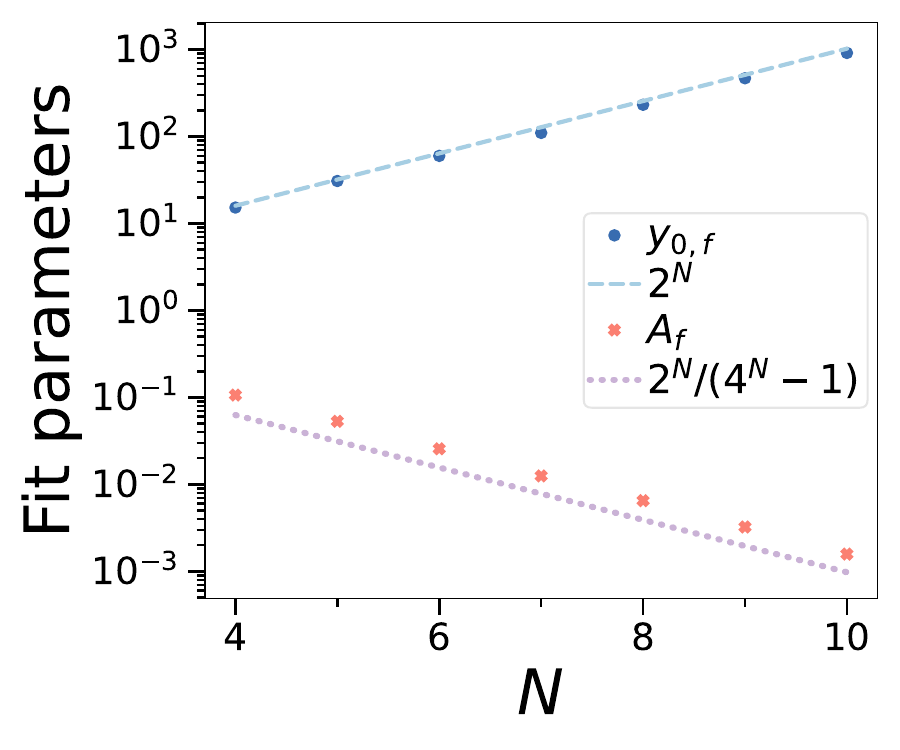}
    \subfigimg[width=0.3\textwidth]{c}{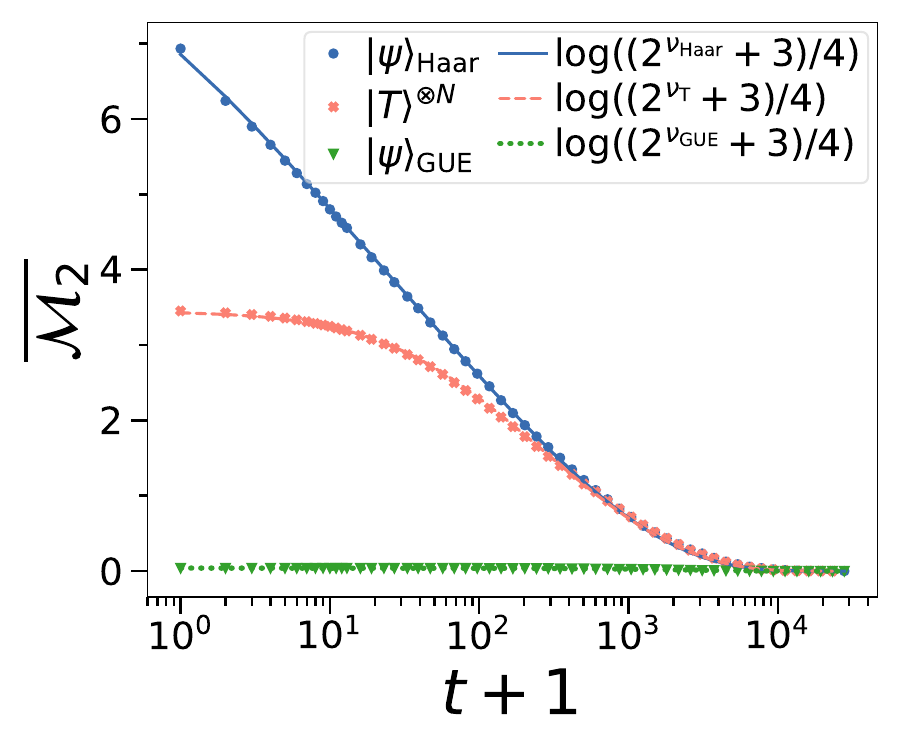}
    \caption{
    \idg{a} Comparison of the average stabilizer nullity $\overline{\nu}$ for different sizes with the approximate differential equation solution in~\eqref{eq:approx_model_nomagic_solution} for different sizes $N$ as a function of measurement steps $t+1$, using $b_f = (1-y_{0,f})/(1+y_{0,f})$ and $A_f$ as fit parameters. 
    \idg{b} Fit parameters $y_{0,f}$ and $A_f$ versus the number of qubits in the system, compared with analytical predictions. In particular, we can see that the predicted value of $y_{0,f} \sim 2^{N}$, which is $2^{\nu(0)}$. The predicted value of $A_N$ in~\eqref{eq:an} is displayed as yellow dashed line. As can be seen from the red dashed line, the fitted value is closer to being $2^{N+1}/(4^N -1)$.
    \idg{c} Average SRE $\overline{\mathcal{M}}_2$ vs time for $\theta_M = 0$ for $\ket{\psi}_{\text{Haar}}$, $\ket{T}^{\otimes N}$ and $\ket{\psi}_{\text{GUE}}$ for a system of $N = 12$ qubits. For all states, dots are the numerical calculations, while for the lines we choose fitting ansatz inspired from our analytics. In particular, thet are the solution of the fitting procedure of~\eqref{eq:fit_m2_nu} with free parameters $y_{0,f}$ and $A_f$.}
    \label{fig:nullity_diff_equation}
\end{figure*}

To further investigate, we can now try and compare also the behavior of the approximate solution in~\eqref{eq:approx_model_nomagic_solution} to the numerical data. In this case, using $A_N$ and $b$ as defined in~\eqref{eq:an} and in~\eqref{eq:b} the general behavior is well represented, but the agreement can be improved, and in order to do that we can use both parameters as a fitting parameters. 
In Fig.~\ref{fig:nullity_diff_equation}(a) we show the numerical data $\overline{\nu}$ as dots, and in dashed lines of the same color the result of the fit using the function and parameters
\begin{equation}
    \overline{\nu}_f = \log_2 (\overline{y}_f(t)) = \log_2 \bigg(1 - \frac{2b_f}{e^{A_ft} + b_f}\bigg), \quad b_f = \frac{1 - y_{0,f}}{1 + y_{0,f}}\,.
    \label{eq:nullity_fit}
\end{equation}
In Fig.~\ref{fig:nullity_diff_equation}(b) we then show, together with the analytical predictions for $A_N$ and $b$, the fitting parameters as a function of the system size $N$. One can see that $y_{0,f}$ as obtained through the fitting procedure is consistent with the analytical prediction $y(0) = 2^N$, (therefore also $b$ is) while the factor $A_f$ as obtained from the fit slightly differs from the analytical prediction in~\eqref{eq:an}, which would explain the not complete agreement between the data and the model.

We can also use the mapping defined in~\eqref{eq:avgsre2_avgnullity} to describe how for $\theta_M = 0$ the SRE decays for initial states different from Haar. This relation is indeed valid for Haar-random states, but it can be extended if the SRE for a general state with maximal initial nullity depends on
\begin{equation}
    \mathcal{M}_2(t) = \ln \bigg( \frac{2^{\nu_f} + 3}{4} \bigg)=\ln \bigg( \frac{2^{\nu_{\text{Haar}}(t)}2^{\Delta \nu(t)} + 3}{4} \bigg)\,,
    \label{eq:fit_m2_nu}
\end{equation}
where $\nu_f$ is the nullity obtained by fitting the SRE of a general initial state, $\nu_{\text{Haar}}(t)$ is the analytic prediction of the nullity and is such that
\begin{equation}
    \nu_f(t) = \nu_{\text{Haar}}(t) + (\nu_f(t) - \nu_{\text{Haar}(t)}) = \nu_{\text{Haar}}(t)  + \Delta \nu(t)\,.
\end{equation}
Defining these quantities, we can finally compare the numerically computes SRE and the one obtained with the fit, and such results are shown in Fig.~\ref{fig:nullity_diff_equation}(c).

\begin{figure}
    \centering
    \subfigimg[width=0.3\textwidth]{}{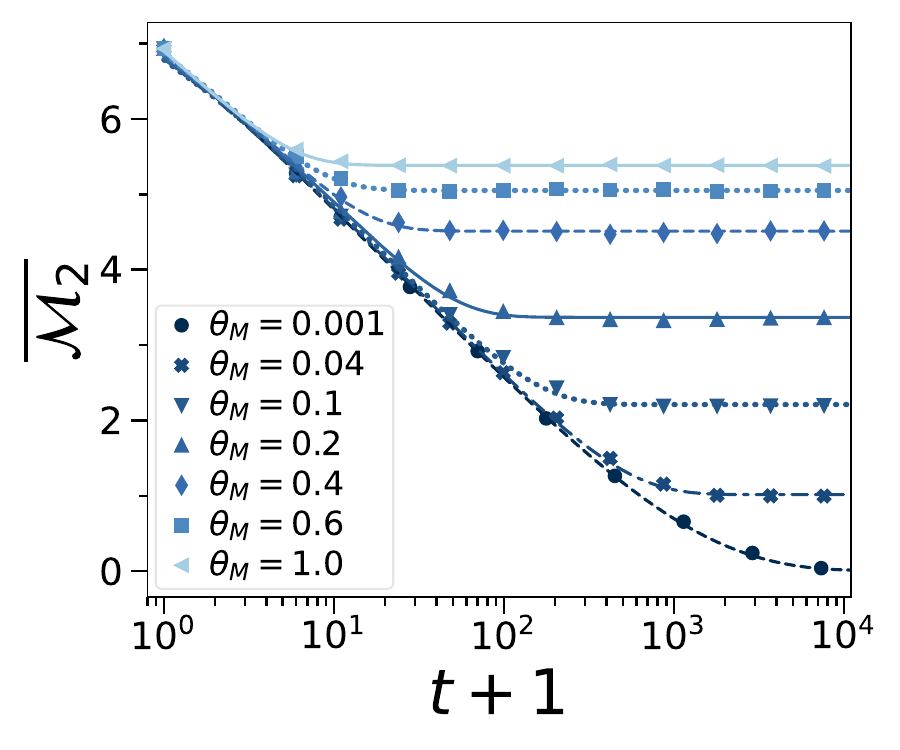}
    \caption{Dots: numerical average SRE for initial $\ket{\psi}_{\text{Haar}}$ versus measurement steps for $N = 12$ qubits for different values of $\theta_M$. Lines: results of the fits performed using~\eqref{eq:gen_fit}.}
    \label{fig:fit_thetam}
\end{figure}

The same procedure can also be extended to the protocol for $\theta_M > 0$ in some cases. Considering that for the protocol in magic basis the $\mathcal{M}_2^{\text{ss}} > 0$, we can also introduce a constant factor in the effective nullity entering SRE as
\begin{equation}
    \label{eq:gen_fit}
    \mathcal{M}_2(\nu) = \ln\bigg(\frac{2^{\nu_f}2^C + 3}{4} \bigg)\,.
\end{equation}
Using now these three free fitting parameters, we are able to describe the behavior of the SRE for Haar-random states for every $\theta_M$, as we show in Fig.~\ref{fig:fit_thetam}. 
In general this fitting model works in cases in which the nullity and the SRE both decrease with the measurement procedure. In the case where the steady-state SRE is larger than the initial one, this model seems to no longer describe the dynamics accurately. 

\section{Deviation from Haar-random behavior}\label{sec:relations_nullity_states}

For random Haar states under random Clifford measurements, we propose in the main text that the state given nullity $\nu$ can be written as
\begin{equation}
\ket{\psi_\nu}=U_\text{C}\ket{0}^{\otimes N-\nu}\ket{M}\,,
\end{equation}
with analytic relationship between the $\alpha$-SRE and the nullity $\nu$ with
\begin{equation}
    \mathcal{M}_2 (\nu) = \ln\bigg(\frac{3 + 2^{\nu}}{4}\bigg)\,.
    \label{eq:sre2_nu}
\end{equation}

\begin{figure*}[htbp]
	\centering	
 	\subfigimg[width=0.3\textwidth]{a}{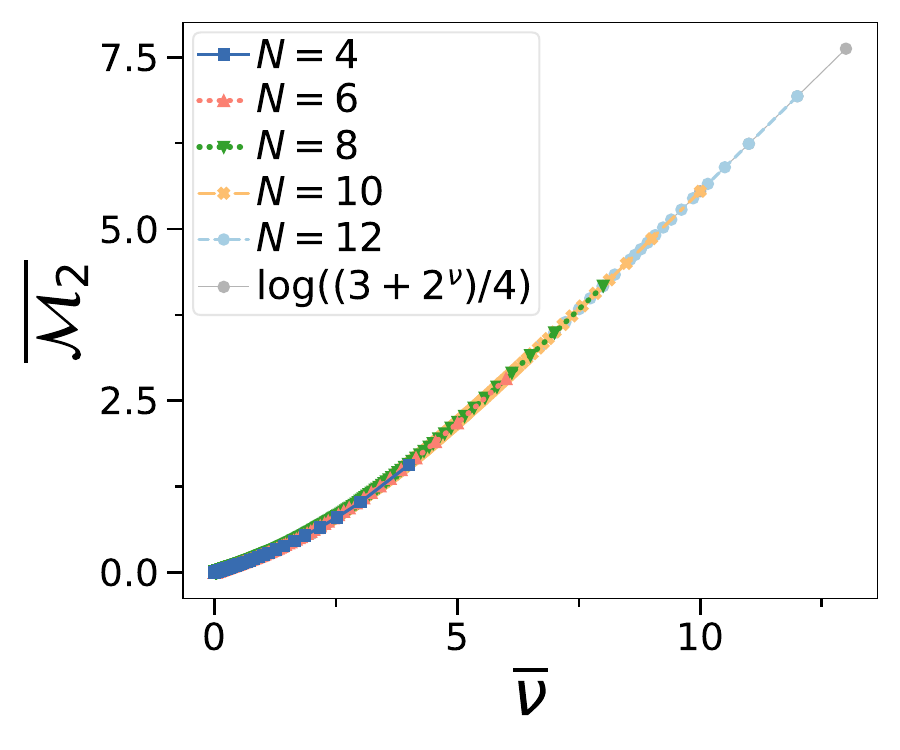}
 	\subfigimg[width=0.3\textwidth]{b}{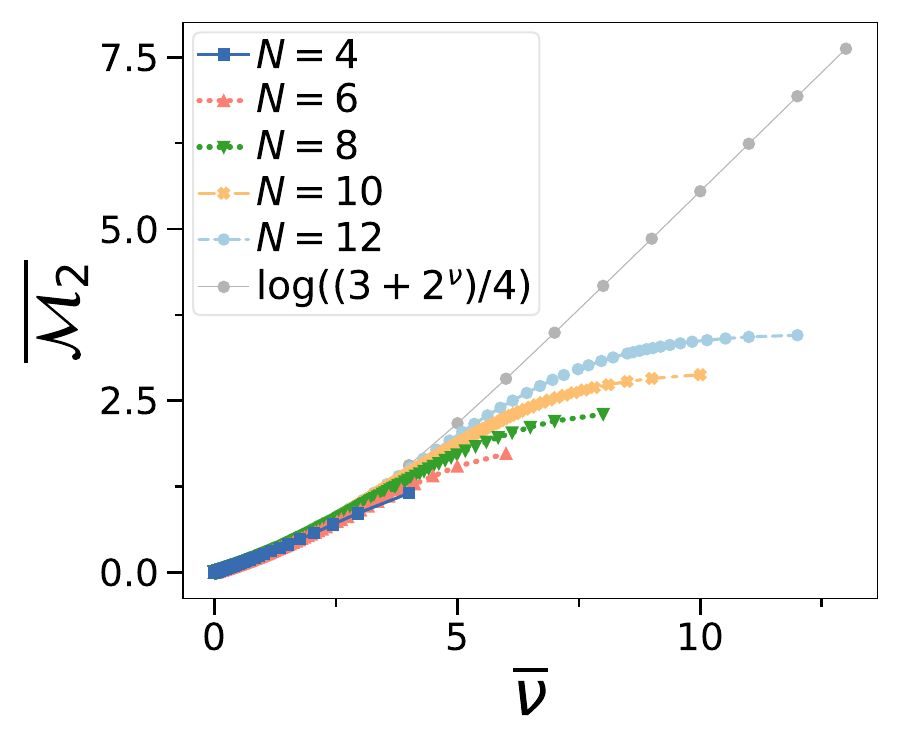}
    \subfigimg[width=0.3\textwidth]{c}{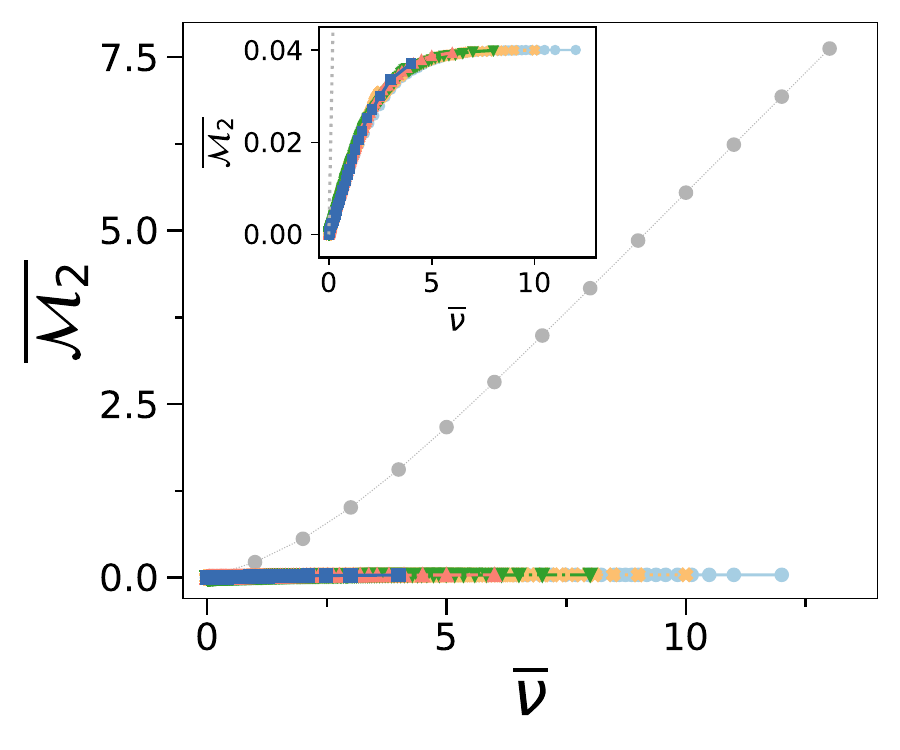}
    \caption{Protocol with $\theta_M = 0$. 
    \idg{a} $\mathcal{M}_2$ against $\nu$ for initial Haar-random states for different $N$, plotted together with the analytic prediction in grey dotted line. 
    \idg{b} $\mathcal{M}_2$ against $\nu$  for initial $\ket{T}^{\otimes N}$ for different $N$, plotted together with the analytic prediction in grey dotted line. 
    \idg{c} $\mathcal{M}_2$ against $\nu$ for initial $\ket{\psi}_\text{GUE}$ state for different $N$. Inset: zoom on main plot. In this case the legend was remove for clarity.}
    \label{fig:sre2_vs_nullity}
\end{figure*}

In Fig.~\ref{fig:sre2_vs_nullity}(a) we plot the average $\mathcal{M}_2$ as a function of the average nullity for random Haar states. We can see that the relation in~\eqref{eq:sre2_nu} is therefore respected at all times, verifying that this protocol sends Haar random states still in Haar random states.

We note that this relationship only works for initial Haar random states. For other types of states such as $\ket{T}^{\otimes N}$ and $\ket{\psi}_\text{GUE}$, we observe distinct dynamics. In particular, we can see that the relation between the two magic quantities for $\ket{T}^{\otimes N}$ as shown in Fig.~\ref{fig:sre2_vs_nullity}b is respected for small values of the nullity, while the magic deviates from the predicted Haar value. Meanwhile, comparing the analytic prediction with the numerical results for $\ket{\psi}_\text{GUE}$, the dynamics if far from the values expected from Haar random states. 

\section{Qubit number dependence of steady-state SREs}
\label{sec:final_value}

\begin{figure}[htbp]
	\centering	
     \subfigimg[width=0.3\textwidth]{}{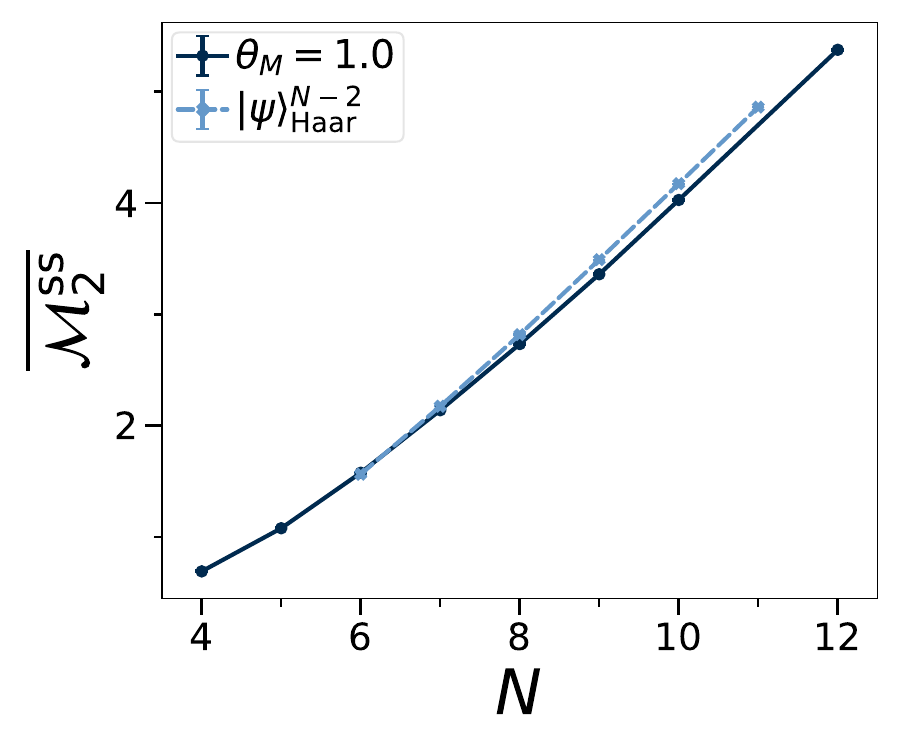}
    \caption{Steady-state SRE  $\overline{\mathcal{M}}_{2}$  against $N$ with $\theta_M=1$ are shown as dots, while the average SRE of Haar-random states of $N-2$ qubits as solid line.
    }
    \label{fig:sre2_theta_gtr0}
\end{figure}

We now study how the steady-state of our circuit depending on qubit number $N$. For $\theta_M = 0$, the steady-state is trivially zero as the final state is a random stabilizer state. 
In particular, in Fig.~\ref{fig:sre2_theta_gtr0} we show the steady-state SRE $\mathcal{M}_{2}^{\text{ss}}$ against $N$ for $\theta_M= 1$.  We find as expected that asymptotically, the SRE increases linearly with $N$. As reference, we also plot the average SRE of Haar-random states for $N - 2$ qubits. Curiously, we find that the steady-state SRE for $N$ qubits nearly matches the SRE of Haar random states of $N-2$. While the steady-state does not have the characteristics of a Haar random state, we leave as an open question whether this relationship between steady-state and Haar random magic has a deeper physical meaning.

\end{document}